\documentclass[prd,floatfix,onecolumn,amsmath,amssymb,floatfix]{revtex4}
\usepackage{graphicx,color,dcolumn,booktabs,bm}
\usepackage{subfigure}
\usepackage{longtable,lscape}
\usepackage{amssymb}
\usepackage{indentfirst}
\usepackage{epsfig}
\usepackage{feynmf}   %{feynmp}
\usepackage{epstopdf}   %{feynmp}
\usepackage{slashed}  %for Feynman symbols
\usepackage{cases}
\usepackage[pdfpages]{xcolor}
\definecolor{maroon}{RGB}{139,25,150}%burada 0-255 arasi her biri icin numara vererek renk elde et
\usepackage{multirow}
\usepackage{float}
\usepackage{graphicx,color,dcolumn,booktabs,bm}
\usepackage[colorlinks, citecolor=blue,anchorcolor=red,menucolor=red, linkcolor=red,filecolor=red,runcolor=red,urlcolor=blue,frenchlinks=red, urlcolor=blue]{hyperref}
\usepackage{orcidlink}

\begin{document}
	\preprint{}
	\preprint{}
	\title{\color{maroon}{Phenomenology of the semileptonic  $\Sigma_{b}^{*0}\,\rightarrow\, \Sigma_{c}^{+}\,\ell\,\bar{\nu}_{\ell}$ transition within QCD sum rules}}

	\author{L.~Khajouei$^{a}$\orcidlink{0000-0003-0181-1879}}
	\email{leila.khajooee@ut.ac.ir}
	\author{K.~Azizi$^{a,b}$\orcidlink{0000-0003-3741-2167}} 
	\email{kazem.azizi@ut.ac.ir} \thanks{Corresponding author} 
	
	\affiliation{
		$^{a}$Department of Physics, University of Tehran, North Karegar Avenue, Tehran 14395-547, Iran\\
		$^{b}$Department of Physics, Dogus University, Dudullu-\"{U}mraniye, 34775 Istanbul, T\"{u}rkiye}

	\date{\today}
	
    \begin{abstract}
	We conduct an investigation on the spin $\frac{3}{2}\rightarrow \frac{1}{2}$ semileptonic weak transition of single heavy baryons for the exclusive decay $\Sigma_{b}^{*0}\,\rightarrow\, \Sigma_{c}^{+}\,\ell\,\bar{\nu}_{\ell}$ in three possible lepton channels within the three point QCD sum rule method. We compute the responsible form factors of this semileptonic decay by  incorporating both perturbative and nonperturbative contributions of the operator product expansion series up to a mass dimension six. Having acquired the form factors, the decay widths of the processes in all lepton channels are determined. Our findings as well as possible future experimental information can be employed in order to check the SM predictions and explore the possibility of new physics in heavy baryonic decay channels.
    \end{abstract}
    \maketitle
    
	\section{Introduction}\label{sec1}
	Probing hadronic structures, as a primary purpose of various experiments and research, can be accomplished by systematic studies of heavy hadrons. Particularly, singly heavy baryons have provoked a great deal of interest. Due to prominent experimental progress, a wide range of these singly heavy baryons have been established in the ground and excited states.   
	   The BABAR Collaboration 
	   \cite{BaBar:2001yhh} 
	   observed the 
	   $\Omega_{c}$
	   \cite{BaBar:2007jdg}  
	   and discovered the 
	   $\Omega_{c}^{*}$
	   state
	    \cite{BaBar:2006pve} in the radiative decay $ \Omega_{c}^{0}\,\gamma$, where the $ \Omega_{c}^{0}$ baryon is reconstructed in the decays to the final states $\Omega^{-}\pi^{+}$ , $\Omega^{-}\pi^{+}\pi^{0}$, $\Omega^{-}\pi^{+}\pi^{-}\pi^{+}$, and $\Xi^{-}K^{-}\pi^{+}\pi^{+}$. 
	    $\Xi_{c}(3055)$ and 
	    $\Xi_{c}(3123)$
	     were observed by BABAR Collaboration
	    \cite{BaBar:2007zjt}
	     and extremely narrow states,\, $\Omega_{c}(3000),\, \Omega_{c}(3050),\, \Omega_{c}(3066), \,\Omega_{c}(3090)$, and 
	     $\Omega_{c}(3119) $ observed by LHCb
	    \cite{LHCb:2017uwr}.
	    The Belle Collaboration found evidence for two excited charm-strange baryon states, $\Xi_{c}(2980)$ and $\Xi_{c}(3077)$, decaying strongly to $\Lambda_{c}^{+}\,K^{0}\,\pi^{-}$ and $\Lambda_{c}^{+}\,K^{-}\,\pi^{+}$ \cite{Belle:2006edu}.
	     In the bottom sector, $\Xi_{b}^{-}$ 
	    \cite{Mizuk:2007ws,D0:2007gjs}
	     was observed by D0
	    \cite{D0:2005cnn}
	     and CDF 
	     \cite{CDF:2004jtw}
	     Collaborations. LHCb established 
	   $\Lambda_{b}(5912)$ and 
	    $\Lambda_{b}(5920)$ in the $\Lambda_{b}^{0}\,\pi^{+}\,\pi^{-}$ spectrum
	    \cite{LHCb:2012kxf}. It also reported the discovery of the 
	    $\Xi_{b}(6327)$ and 
	    $\Xi_{b}(6333)$
	    \cite{LHCb:2021ssn}. The CDF Collaboration presented an observation of four $\Lambda_{b}^{0}\,\pi^{\pm}$ resonances interpreted as 
	    $\Sigma_{b}^{\pm}$ and 
	    $\Sigma_{b}^{*\pm}$ baryons in the reconstructed decay mode $\Lambda_{b}^{0}\,\rightarrow \,\Lambda_{c}^{+}\,\pi^{-}$, where $\Lambda_{c}^{+}\,\rightarrow\,pK^{-}\pi^{+}$ \cite{CDF:2007oeq}. 
	    
	    The study of singly heavy baryons from different aspects, especially decay properties, is of crucial importance. A variety of theoretical investigations into the strong, weak, and radiative decays of the singly heavy baryons can be found in the  literature\cite{Aliev:2009jt,Olamaei:2021eyo,Garcia-Tecocoatzi:2023btk,Yao:2018jmc,Wang:2017kfr,Efimov:1991qj,Ortiz-Pacheco:2023kjn,Ivanov:1996fj,Aliev:2010yx,Aliev:2008sk,Azizi:2011if}. Weak decays of singly heavy baryons receive great concentration since a detailed experimental survey on various weak decay channels can be employed to obtain the standard model (SM) parameters, such as the CKM matrix elements. Any obvious deviation from SM predictions, clarify the feasible existence of new physics effects beyond the standard model (BSM). In particular, the semileptonic weak decays provide us with a lot of useful information, and they have been discussed within several approaches such as the 
	    relativistic quark model \cite{Ebert:2006rp}, 
	    covariant confined quark model \cite{Gutsche:2015mxa},
	    light cone QCD sum rules \cite{Aliev:2010uy,Aliev:2023tpk,Azizi:2011mw,Wang:2009yma}, 
	    nonrelativistic quark model \cite{Cheng:1995fe}, 
	    light front quark model \cite{Ke:2019smy}, 
	    and lattice QCD \cite{Detmold:2015aaa}. Most of these studies concentrate on spin 
	    $\frac{1}{2}\rightarrow \frac{1}{2}$ and 
	    $\frac{1}{2}\rightarrow \frac{3}{2}$ transitions, which the total angular momentum of the initial singly heavy baryon is 
	    $\frac{1}{2}$.

	    In this work, we study the semileptonic weak decay of a singly heavy baryon with total angular momentum 
	    $J=\frac{3}{2}$, $\Sigma_{b}^{*0}$, into a singly heavy baryon with 
	    $J=\frac{1}{2}$, $\Sigma_{c}^{+}$,  
	    within the three point QCD sum rule method. This approach is a powerful and predictive nonperturbative method in which a hadron-to-hadron transition matrix element is expressed in terms of a correlation function in both hadronic and quark-gluon level. Considering the previous consistent predictions of this method with the experiments, it is found to be a reliable approach and can be employed to determine different hadronic parameters \cite{Agaev:2017jyt,Agaev:2017lip,Aliev:2010uy,Agaev:2016dev,Azizi:2018axf,Azizi:2016dhy,Neishabouri:2024gbc}. In particular, we consider the 
	    $b\rightarrow c$ transition at quark level in semileptonic weak decay 
	    $\Sigma_{b}^{*0}\rightarrow \Sigma_{c}^{+} \ell \bar{\nu}_{\ell}$ and compute the responsible form factors and decay widths in three lepton channels.  
	    
	    The structure of this paper is organized as follows. In Sec. \ref{sec sum}, the construction of the QCD sum rules for corresponding form factors of the $\Sigma_{b}^{*0}\rightarrow \Sigma_{c}^{+}\,\ell\,\bar{\nu}_{\ell}$ transition is explained. Numerical results after analyzing the obtained sum rules for transition form factors and fixing the working regions of auxiliary parameters, are exhibited in Sec. \ref{analysis}. Utilizing the fit functions of form factors in terms of transferred momentum squared, the decay widths of these processes in all lepton channels are calculated in Sec. \ref{width}. Finally, Sec. \ref{clue} is dedicated to the Summary and Conclusion; and some features of the computations are presented in the Appendixes.

	   \section{The method}\label{sec sum}           
	 \subsection{Weak transition form factors}\label{ff}
	 The semileptonic weak transitions originate from a quark weak current, which is connected with a leptonic weak current via a W boson. Due to the large W-boson's mass, an effective lepton-quark interaction may occur, and it can be described by an effective Hamiltonian.
	
	In the present study, we focus on the semileptonic weak decay of the 
	$\Sigma_{b}^{*0}$ that is a single heavy baryon with one b heavy quark (See Table I). 
	 \begin{table}[h]
		\centering
		\begin{tabular}{ccccc}
			\hline
			Baryon\,\,\,&Quark content\,\,\,&Charge\,\,\,&Quark model\,\,\,&Spin parity\,\\
			\hline
			$\Sigma_{b}^{*0}$&$(u\,d\,b)$&0&sextet&$\frac{3}{2}^{+}$\\
			\hline
			\hline
		\end{tabular}
		\caption{ 
			Quantum numbers and quark content of\,\,$\Sigma_{b}^{*0}$.}
	\end{table}

	We consider the 
	$b\rightarrow c\, \ell \,\bar{\nu}_{\ell}$ transition in 
	$\Sigma_{b}^{*0}\rightarrow \Sigma_{c}^{+}\, \ell \,\bar{\nu}_{\ell}$ decay channel which the 
	$\Sigma_{c}^{+}$ is also a single heavy baryon with one c heavy quark (See Table II).
	 \begin{table}[h]
		\centering
		\begin{tabular}{ccccc}
			\hline
			Baryon\,\,\,&Quark content\,\,\,&Charge\,\,\,&Quark model\,\,\,&Spin parity\,\\
			\hline
			$\Sigma_{c}^{+}$&$(u\,d\,c)$&+1&sextet&$\frac{1}{2}^{+}$\\
			\hline
			\hline
		\end{tabular}
		\caption{ 
			Quantum numbers and quark content of\,\, $\Sigma_{c}^{+}$.}
	\end{table}
	
	An effective quark-lepton interaction, as the source of
	$b\rightarrow c\, \ell \,\bar{\nu}_{\ell}$ transition, can be described by the Hamiltonian,
	 \begin{equation}
		\mathcal{H}_{eff}\,=\,\frac{G_{F}}{\sqrt{2}}\,V_{cb}\,\bar{c}\,\gamma_{\mu}\,(\,1-\gamma_{5}\,)\,b\,\bar{\ell}\,\gamma^{\mu}\,(\,1-\gamma_{5}\,)\,\nu_{\ell},
	\end{equation} 
	where 
	 $G_{F}$
	 is the Fermi coupling constant and 
	  $V_{cb}$
	  is one of the elements of the CKM matrix which parametrizes the quark-flavor mixing in SM.
	  By putting this effective Hamiltonian between the initial and final baryonic states, the amplitude of this decay mode is found,
	   \begin{equation}
	  	M\,=\,\langle\,\Sigma_{c}^{+}|\,\mathcal{H}_{eff}\,|\,\Sigma_{b}^{*0}\,\rangle,
	  \end{equation}
    where the pointlike leptonic parts, go out of the matrix element and the other parts are parametrized, considering the Lorentz invariance and parity requirements, in terms of eight form factors;  four for vector current, $J_{\mu}^{V}$, which are 
     $F_{1}(q^{2})$,  
    $F_{2}(q^{2})$, 
    $F_{3}(q^{2})$, 
    $F_{4}(q^{2})$,
     and the other four for axial-vector current, $J_{\mu}^{A}$, which are
    $G_{1}(q^{2})$,  
    $G_{2}(q^{2})$, 
    $G_{3}(q^{2})$,
     and
    $G_{4}(q^{2})$,
    \begin{eqnarray}\label{formfactor}
    			&&\langle\,\Sigma_{c}^{+}(p^{\prime},s^{\prime})\,|\,J_{\mu}^{V}\,|\,\Sigma_{b}^{*0}(p,s)\,\rangle\,=\,
    			\langle\,\Sigma_{c}^{+}(p^{\prime},s^{\prime})\,|\,\bar{c}\,\gamma_{\mu}\,b\,|\,\Sigma_{b}^{*0}(p,s)\,\rangle\,\notag\\
    			&&=\,\bar{u}_{\Sigma_{c}^{+}}(p^{\prime},s^{\prime})\,\big[g_{\mu\alpha}\,F_{1}(q^{2})+\gamma_{\mu}\,\frac{p^{\prime}_{\alpha}}{M_{\Sigma_{c}^{+}}}\,F_{2}(q^{2})+\frac{p^{\prime}_{\alpha}\,p_{\mu}}{M_{\Sigma_{c}^{+}}^{2}}\,F_{3}(q^{2})+\frac{p^{\prime}_{\alpha}\,q_{\mu}}{M_{\Sigma_{c}^{+}}^{2}}\,F_{4}(q^{2})\big]\,\gamma_{5}\,u_{\Sigma_{b}^{*0}}^{\alpha}(p,s),\notag\\
    			&&\notag\\
    			&&\langle\,\Sigma_{c}^{+}(p^{\prime},s^{\prime})\,|\,J_{\mu}^{A}\,|\,\Sigma_{b}^{*0}(p,s)\,\rangle\,=\,
    			\langle\,\Sigma_{c}^{+}(p^{\prime},s^{\prime})\,|\,\bar{c}\,\gamma_{\mu}\,\gamma_{5}\,b\,|\,\Sigma_{b}^{*0}(p,s)\,\rangle\,\notag\\
    			&&=\,\bar{u}_{\Sigma_{c}^{+}}(p^{\prime},s^{\prime})\,\big[g_{\mu\alpha}\,G_{1}(q^{2})+\gamma_{\mu}\,\frac{p^{\prime}_{\alpha}}{M_{\Sigma_{c}^{+}}}\,G_{2}(q^{2})+\frac{p^{\prime}_{\alpha}\,p_{\mu}}{M_{\Sigma_{c}^{+}}^{2}}\,G_{3}(q^{2})+\frac{p^{\prime}_{\alpha}\,q_{\mu}}{M_{\Sigma_{c}^{+}}^{2}}\,G_{4}(q^{2})\big]\,u_{\Sigma_{b}^{*0}}^{\alpha}(p,s),
    \end{eqnarray}	
    where 
     $ q\,=\,p\,-\,p^{\prime} $
     is the momentum transferred to the lepton and corresponding antineutrino. 
     $u_{\Sigma_{c}^{+}}(p^{\prime}, s^{\prime})$ and
      $u_{\Sigma_{b}^{*0}}^{\alpha}(p, s)$ are Dirac spinor of the final baryonic state with momentum $p^{\prime}$ and spin $s^{\prime}$ and Rarita-Schwinger spinor of the initial state with momentum $p$ and spin $s$, respectively.
      Form factors are building blocks of the transitions under study, and they are Lorentz invariant. 
      
      The principal aim of this work is to compute these transition form factors via the three-point QCD sum rules method \cite{Shifman:2010zzb, Shifman:1978by, Shifman:1978bx, Gross:2022hyw}. Being based on the QCD Lagrangian is the interest of this nonperturbative framework. In this approach, a baryonic transition is described by a correlation function incorporating the initial and final baryonic currents connected with a transition one.  
      
     We consider the following three-point correlation function: 
      \begin{equation}\label{correlation4}
     	\Pi_{\mu\nu}(p,p^{\prime},q^2)\,=\,i^{2}\,\int\,\mathrm{d}^{4}x\,e^{-ip.x}\,\int\,\mathrm{d}^{4}y\,e^{ip^{\prime}.y}\,\langle\,0\,|\,{\cal T}\,\big\lbrace {\cal J}^{\Sigma_{c}^{+}}(y)\,{\cal J}_{\mu}^{tr}(0)\,\bar{\cal J}_{\nu}^{\Sigma_{b}^{*0}}(x)\big\rbrace\,|\,0\,\rangle,
     \end{equation} 
      where ${\cal T}$ stands for the time-ordering operator and 
      ${\cal J}^{\Sigma_{b}^{*0}}$ and 
      ${\cal J}^{\Sigma_{c}^{+}}$ are the interpolating currents for the initial and final baryons, respectively, and ${\cal J}_{\mu}^{tr}$ is the transition current. The correlation function is a function of 
      $q^2$. If $q^2$ is shifted from spacelike to timelike region, the long-distance quark-gloun interactions become dominant and hadrons are formed. In this framework, the correlation function is computed first in hadronic level, which is called phenomenological or physical side and then in quark-gluon level, using the operator product expansion  (OPE), that is named theoretical or QCD side. These two representations are then matched to give QCD sum rules for the form factors.  
      \subsection{Phenomenological side}  
    The correlation function receives contributions of the ground state, excited, and a continuum of many-body hadron states. A proper way to quantify the complex hadronic content of the correlation function is provided by the unitarity relation that is obtained by inserting  complete sets of the hadronic states with the same quantum numbers as the interpolating currents of the initial and final hadrons into the correlation function \cite{Khodjamirian:2020btr,Colangelo:2000dp},  
     \begin{equation}
    	1\,=\,|\,0\rangle\,\langle0\,|+\sum_{h}\,\int\,\frac{\mathrm{d}^{4}p_{h}}{(2\pi)^{4}}\,2\pi\,\delta(p_{h}^{2}-m_{h}^{2})|\,h(p_{h})\,\rangle\,\langle\,h(p_{h})\,|\,\,+\,\text{higher Fock states},
    \end{equation}
    where $|\,h(p_{h})\rangle$ denotes the possible hadronic state with momentum $p_{h}$.
    
    After inserting the complete sets in the correlation function, Eq. \eqref{correlation4}, and performing the Fourier transformation by integrating over four-x and -y, we obtain
      \begin{equation}\label{physical}
    	\Pi_{\mu\nu}^{Phys.}(p,p^{\prime},q^{2})\,=\,\frac{\langle\,0\,|\,{\cal J}^{\Sigma_{c}^{+}}(0)\,|\,\Sigma_{c}^{+}(p^{\prime})\,\rangle\,\langle\,\Sigma_{c}^{+}(p^{\prime})|{\cal J}_{\mu}^{tr}(0)|\Sigma_{b}^{*0}(p)\,\rangle\,\langle\,\Sigma_{b}^{*0}(p)|\bar{\cal J}_{\nu}^{\Sigma_{b}^{*0}}(0)\,|\,0\rangle}{(p^{\prime2}-m_{\Sigma_{c}^{+}}^{2})(p^{2}-m_{\Sigma_{b}^{*0}}^{2})}\,+\,\cdots ,
    \end{equation}   
   where the $\cdots$ denote the higher states and continuum.
   
   The matrix element $\langle\,0\,|\,{\cal J}^{\Sigma_{c}^{+}}(0)\,|\,\Sigma_{c}^{+}(p^{\prime})\,\rangle$ is defined in terms of the residue of the final baryonic state, $\lambda^{\prime}_{\Sigma_{c}^{+}}$, as,
    \begin{equation}\label{resc}
   	\langle\,0\,|\,{\cal J}^{\Sigma_{c}^{+}}(0)\,|\,\Sigma_{c}^{+}(p^{\prime})\,\rangle\,=\,\lambda^{\prime}_{\Sigma_{c}^{+}}\,u_{\Sigma_{c}^{+}}(p^{\prime},s^{\prime}),
   \end{equation}
     and $\langle\,\Sigma_{b}^{*0}(p)\,|\,\bar{\cal J}_{\nu\Sigma_{b}^{*0}}(0)\,|\,0\,\rangle$ can be parametrized in terms of the residue of the initial state, $\lambda_{\Sigma_{b}^{*0}}$,
      \begin{equation}\label{resb}
     	\langle\,\Sigma_{b}^{*0}(p)\,|\,\bar{\cal J}_{\nu}^{\Sigma_{b}^{*0}}(0)\,|\,0\,\rangle\,=\,\lambda_{\Sigma_{b}^{*0}}^{\dagger}\,\bar{u}_{\nu\Sigma_{b}^{*0}}(p,s),
     \end{equation}  
    
    After substituting Eqs. \eqref{formfactor}, \eqref{resc}, and \eqref{resb} in Eq. \eqref{physical}, since the initial and final baryons are unpolarized, we utilize the summation relations over Dirac spinors for final baryon with spin $\frac{1}{2}$, 
     \begin{equation}
    	\sum_{s^{\prime}}\,\,u_{\Sigma_{c}^{+}}(p^{\prime},s^{\prime})\,\bar{u}_{\Sigma_{c}^{+}}(p^{\prime},s^{\prime})\,=\,(\slashed{p}^{\prime}\,+\,m_{\Sigma_{c}^{+}}),
    \end{equation}  
     and Rarita-Schwinger spinors for initial baryon with spin $\frac{3}{2}$,
      \begin{equation}
     	\sum_{s}\,\,u_{\alpha\,\Sigma_{b}^{*0}}(p,s)\,\bar{u}_{\nu\,\Sigma_{b}^{*0}}(p,s)\,=\,-(\slashed{p}\,+\,m_{\Sigma_{b}^{*0}})\,\Big(\,g_{\alpha\nu}\,-\,\frac{\gamma_{\alpha}\gamma_{\nu}}{3}\,-\,\frac{2}{3}\,\frac{p_{\alpha}p_{\nu}}{m_{\Sigma_{b}^{*0}}^{2}}\,+\,\frac{1}{3}\,\frac{p_{\alpha}\gamma_{\nu}-p_{\nu}\gamma_{\alpha}}{m_{\Sigma_{b}^{*0}}}\Big),
     \end{equation}
      to obtain the physical side of the correlation function.
      
      Before proceeding with more computations, we make the following remarks. First, the interpolating current of the spin $\frac{3}{2}$ initial baryon, couples not only to spin $\frac{3}{2}$ state, but also to spin $\frac{1}{2}$ state. Imposing the condition  $\gamma^{\nu}{\cal J}_{\nu}\,=\,0$, these contributions can be written as
     
      \begin{equation} \langle0\,|{\cal J}_{\nu}^{\Sigma_{b}^{*0}}|\,(p,s=\frac{1}{2})\rangle\,=\,(\alpha\gamma_{\nu}\,-\frac{4\alpha}{m} p_{\nu})\,u(p,s=\frac{1}{2}).
      \end{equation}  
      It expresses that the Lorentz structures $\gamma_{\nu}$ and $p_{\nu}$ have contributions of spin-$\frac{1}{2}$ baryons. Hence, the structures proportional to $\gamma_{\nu}$ at the far right end or $p_{\nu}$ must be eliminated properly. Second, the structures, which appear in the phenomenological or physical side of the correlation function, are not all independent, and it is essential to order the matrices in a specific form. In this work, the matrices are ordered in the form $\gamma_{\mu}\slashed{p}\slashed {p}^{\prime}\gamma_{\nu}\gamma_{5}$. 
      
      In the three-point QCD sum rule method, there are three kinds of auxiliary parameters, the Borel mass parameters ($M^2,\, M^{\prime 2}$) and the continuum thresholds ($s_0,\, s^{\prime}_0$) for initial and final baryonic channels, respectively which have been brought by applying the Borel transformation and the continuum subtraction to suppress the contributions coming from the higher states and continuum, and the third one is $\beta$, a mathematical mixing parameter inside the  current of  spin $\frac{1}{2}$ state. The working regions of these parameters are fixed by requiring that the physical observables  be possibly  independent of them  as well as making use of  other related prescriptions that will be discussed later.
      
      We apply the double Borel transformation to the physical side using \cite{Aliev:2006gk}
      \begin{equation}\label{Borel} 
      	\mathbf{\widehat{B}}\,\frac{1}{(p^2-m^{2})^{m}}\,\frac{1}{(p^{\prime2}-m^{\prime2})^{n}}\,\longrightarrow\,(-1)^{m+n}\,\frac{1}{\Gamma[m]\,\Gamma[n]}\,\frac{1}{(M^{2})^{m-1}}\,\frac{1}{(M^{\prime2})^{n-1}}\,e^{-m^{2}/M^{2}}\,e^{-m^{\prime2}/M^{\prime2}},
      \end{equation}	 
      where $M^2$ and $M^{\prime 2}$ are Borel mass parameters and $m$ and $m^{\prime}$ are the masses of the initial and final baryons. 
      
      Utilization of all mentioned above leads us to the final form of the phenomenological side of the correlation function as
       \begin{eqnarray}\label{physical1}
      		&&\mathbf{\widehat{B}}\,\Pi_{\mu\nu}^{Phys.}(p,p^{\prime},q^2)\,=\,\lambda_{\Sigma_{b}^{*0}}\,\lambda_{\Sigma_{c}^{+}}\,e^{-\frac{m_{\Sigma_{b}^{*0}}^{2}}{M^{2}}}\,e^{-\frac{m_{\Sigma_{c}^{+}}^{2}}{M\prime^{2}}}\,\Bigg[F_{1}\Big(-m_{\Sigma_{b}^{*0}}\,m_{\Sigma_{c}^{+}}\,g_{\mu\nu}\,\gamma_{5}+m_{\Sigma_{c}^{+}}\,g_{\mu\nu}\,\slashed{p}\,\gamma_{5}-m_{\Sigma_{b}^{*0}}\,g_{\mu\nu}\,\slashed{p}^{\prime}\,\gamma_{5}-g_{\mu\nu}\,\slashed{p}\,\slashed{p}^{\prime}\,\gamma_{5}\Big)+\notag\\&&F_{2}\Big(-m_{\Sigma_{b}^{*0}}\,p^{\prime}_{\nu}\,\gamma_{\mu}\,\gamma_{5}+p^{\prime}_{\nu}\,\gamma_{\mu}\,\slashed{p}\,\gamma_{5}+\frac{m_{\Sigma_{b}^{*0}}}{m_{\Sigma_{c}^{+}}}\,p^{\prime}_{\nu}\,\gamma_{\mu}\,\slashed{p}^{\prime}\,\gamma_{5}+\frac{1}{m_{\Sigma_{c}^{+}}}\,p^{\prime}_{\nu}\,\gamma_{\mu}\,\slashed{p}\,\slashed{p}^{\prime}\,\gamma_{5}\Big)+F_{3}\Big(-\frac{m_{\Sigma_{b}^{*0}}}{m_{\Sigma_{c}^{+}}}\,p_{\mu}\,p^{\prime}_{\nu}\,\gamma_{5}+\frac{1}{m_{\Sigma_{c}^{+}}}\,p_{\mu}\,p^{\prime}_{\nu}\,\slashed{p}\,\gamma_{5}-\notag\\&&\frac{m_{\Sigma_{b}^{*0}}}{m_{\Sigma_{c}^{+}}^{2}}\,p_{\mu}\,p^{\prime}_{\nu}\slashed{p}^{\prime}\,\gamma_{5}-\frac{1}{m_{\Sigma_{c}^{+}}^{2}}\,p_{\mu}\,p^{\prime}_{\nu}\,\slashed{p}\,\slashed{p}^{\prime}\,\gamma_{5}\Big)+F_{4}\Big(\frac{m_{\Sigma_{b}^{*0}}}{m_{\Sigma_{c}^{+}}^{2}}\,p^{\prime}_{\mu}\,p^{\prime}_{\nu}\,\slashed{p}^{\prime}\,\gamma_{5}-\frac{m_{\Sigma_{b}^{*0}}}{m_{\Sigma_{c}^{+}}^{2}}\,p_{\mu}\,p^{\prime}_{\nu}\,\slashed{p}^{\prime}\,\gamma_{5}-\frac{1}{m_{\Sigma_{c}^{+}}^{2}}\,p_{\mu}\,p^{\prime}_{\nu}\,\slashed{p}\,\slashed{p}^{\prime}\,\gamma_{5}+\frac{1}{m_{\Sigma_{c}^{+}}^{2}}\,p^{\prime}_{\mu}\,p^{\prime}_{\nu}\,\slashed{p}\,\slashed{p}^{\prime}\,\gamma_{5}\Big)\notag\\&&-G_{1}\Big(-m_{\Sigma_{b}^{*0}}\,m_{\Sigma_{c}^{+}}\,g_{\mu\nu}\,-m_{\Sigma_{c}^{+}}\,g_{\mu\nu}\,\slashed{p}\,-m_{\Sigma_{b}^{*0}}\,g_{\mu\nu}\,\slashed{p}^{\prime}\,+g_{\mu\nu}\,\slashed{p}\,\slashed{p}^{\prime}\,\Big)-G_{2}\Big(-m_{\Sigma_{b}^{*0}}\,p^{\prime}_{\nu}\,\gamma_{\mu}\,-p^{\prime}_{\nu}\,\gamma_{\mu}\,\slashed{p}\,+\frac{m_{\Sigma_{b}^{*0}}}{m_{\Sigma_{c}^{+}}}\,p^{\prime}_{\nu}\,\gamma_{\mu}\,\slashed{p}^{\prime}\,\notag\\&&-\frac{1}{m_{\Sigma_{c}^{+}}}\,p^{\prime}_{\nu}\,\gamma_{\mu}\,\slashed{p}\,\slashed{p}^{\prime}\Big)\,-G_{3}\Big(-\frac{m_{\Sigma_{b}^{*0}}}{m_{\Sigma_{c}^{+}}}\,p_{\mu}\,p^{\prime}_{\nu}\,-\frac{1}{m_{\Sigma_{c}^{+}}}\,p_{\mu}\,p^{\prime}_{\nu}\,\slashed{p}\,-\frac{m_{\Sigma_{b}^{*0}}}{m_{\Sigma_{c}^{+}}^{2}}\,p_{\mu}\,p^{\prime}_{\nu}\slashed{p}^{\prime}\,+\frac{1}{m_{\Sigma_{c}^{+}}^{2}}\,p_{\mu}\,p^{\prime}_{\nu}\,\slashed{p}\,\slashed{p}^{\prime}\,\Big)-G_{4}\Big(\frac{m_{\Sigma_{b}^{*0}}}{m_{\Sigma_{c}^{+}}^{2}}\,p^{\prime}_{\mu}\,p^{\prime}_{\nu}\,\slashed{p}^{\prime}\,\notag\\&&-\frac{m_{\Sigma_{b}^{*0}}}{m_{\Sigma_{c}^{+}}^{2}}\,p_{\mu}\,p^{\prime}_{\nu}\,\slashed{p}^{\prime}\,+\frac{1}{m_{\Sigma_{c}^{+}}^{2}}\,p_{\mu}\,p^{\prime}_{\nu}\,\slashed{p}\,\slashed{p}^{\prime}\,-\frac{1}{m_{\Sigma_{c}^{+}}^{2}}\,p^{\prime}_{\mu}\,p^{\prime}_{\nu}\,\slashed{p}\,\slashed{p}^{\prime}\,\Big)\Bigg]+\cdots,
      \end{eqnarray}
      where we omitted the $ q^2 $ dependence of form factors for simplicity.
     \subsection{Theoretical side} 
     As mentioned in Sec. \ref{ff}, the correlation function is also defined at negative values of $q^{2}$. In a deep Euclidean region,  $Q^{2}\,=\,-q^{2}\,\gg\,\Lambda_{QCD}^{2}$, the correlation function can be expressed in terms of the quark and gluon fields, considering the QCD - vacuum fluctuations and both perturbative and nonperturbative effects, and computed using the OPE,
      \begin{equation}
     	\Pi(q^{2})\,=\,\sum_{d}\,C_{d}(q^{2})\,\langle0\,|O_{d}|\,0\rangle,
     \end{equation}
     where $C_{d}(q^{2})$ are Wilson coefficients and the operators $O_{d}$ are arranged according to their dimension $d$ and the vacuum expectation values of these operators are denoted in a variety of $vacuum \,\,condensates$.
     Thus the next step, in the three-point QCD sum rules method, is to compute the theoretical or QCD‌ side of the correlation function. For this goal, the explicit forms of the interpolating currents of the initial and final baryons are embedded into the correlation function, Eq. \eqref{correlation4}. These currents are expressed by the corresponding quark fields and given as  
      \begin{equation}\label{currentb}
     		{\cal J}_{\mu}^{\Sigma_{b}^{*0}}(x)\,=\,\sqrt\frac{2}{3}\,\epsilon^{abc}\Bigg\lbrace \Big(u^{aT}(x)C\gamma_{\mu}\,d^{b}(x)\,\Big)b^{c}(x)\,+\,\Big(d^{aT}(x)C\gamma_{\mu}\,b^{b}(x)\,\Big)u^{c}(x)\,+\,\Big(b^{aT}(x)C\gamma_{\mu}\,u^{b}(x)\,\Big)d^{c}(x)\,\Bigg\rbrace,
     \end{equation}   
     for the initial singly heavy baryon with spin $\frac{3}{2}$ (for details see the Appendix  \ref{appA}) and 
      \begin{eqnarray}\label{currentc}
     		{\cal J}^{\Sigma_{c}^{+}}(y)\,=\,-\frac{1}{\sqrt{2}}\,\epsilon^{abc}\,\Bigg\lbrace \Big(u^{aT}(y)C\,c^{b}(y)\,\Big)\gamma_{5}d^{c}(y)\,+\,\beta\,\Big(u^{aT}(y)C\gamma_{5}\,c^{b}(y)\,\Big)d^{c}(y)\,\nonumber\\
     		-\,\Big(c^{aT}(y)C\,d^{b}(y)\,\Big)\gamma_{5}\,u^{c}(y)-\beta\,\Big(c^{aT}(y)\,C\,\gamma_{5}\,d^{b}(y)\Big)\,u^{c}(y)\,\Bigg\rbrace,
     \end{eqnarray}
      for the final singly heavy baryon with spin $\frac{1}{2}$, where  $a$, $b$, and $c$ are color indices, C is the charge conjugation operator, and $u(x)$, $d(x)$,  $c(x)$, and  $b(x)$ are quark fields. After substituting Eqs. \eqref{currentb} and \eqref{currentc} in Eq.\eqref{correlation4} and contracting the related quark fields, we acquire the QCD side of the correlation function in terms of the heavy and light quark propagators in coordinate space in the following form:  
       \begin{eqnarray}\label{qcdside}
      		&&\Pi_{\mu\nu}^{QCD}(p,p^{\prime },q^{2})=\,i^{2} \int\,\mathrm{d}^{4}x\,e^{-ipx}\,\int\,\mathrm{d}^{4}y\,e^{ip^{\prime}y}\,\frac{1}{\sqrt{3}}\,\epsilon^{abc}\,\epsilon^{a^{\prime} b^{\prime} c^{\prime}}\,\nonumber\\&&\Bigg\lbrace\,\gamma_{5}\,S_{d}^{cb^{\prime}}(y-x)\,\gamma_{\nu}\,\overset{\sim}{S}_{u}^{aa^{\prime}}(y-x)\,S_{c}^{bi}(y)\,\big(\gamma_{\mu}\,(1-\gamma_{5})\big)\,S_{b}^{ic^{\prime}}(-x)\nonumber\\
      		&&-\gamma_{5}\,S_{d}^{ca^{\prime}}(y-x)\,\gamma_{\nu}\,\overset{\sim}{S}_{b}^{ib^{\prime}}(-x)\,\big(\,(1-\gamma_{5})\gamma_{\mu}\big)\,\overset{\sim}{S}_{c}^{bi}(y)S_{u}^{ac^{\prime}}(y-x)\nonumber\\
      		&&-\gamma_{5}\,S_{d}^{cc^{\prime}}(y-x)\,Tr[\,S_{u}^{ab^{\prime}}(y-x)\gamma_{\nu}\,\overset{\sim}{S}_{b}^{ia^{\prime}}(-x)\,\big(1-\gamma_{5})\gamma_{\mu}\big)\,\overset {\sim}{S}_{c}^{bi}(y)]\nonumber\\
      		&&+\beta\,S_{d}^{cb^{\prime}}(y-x)\,\gamma_{\nu}\,\overset{\sim}{S}_{u}^{aa^{\prime}}(y-x)\gamma_{5}S_{c}^{bi}(y)\,\big(\gamma_{\mu}\,(1-\gamma_{5})\big)\,S_{b}^{ic^{\prime}}(-x)\nonumber\\
      		&&-\beta\,S_{d}^{ca^{\prime}}(y-x)\,\gamma_{\nu}\,\overset{\sim}{S}_{b}^{ib^{\prime}}(-x)\,\big(\,(1-\gamma_{5})\gamma_{\mu}\big)\,\overset{\sim}{S}_{c}^{bi}(y)\gamma_{5}S_{u}^{ac^{\prime}}(y-x)\nonumber\\
      		&&-\beta\,S_{d}^{cc^{\prime}}(y-x)Tr[\,\gamma_{\nu}\,\overset{\sim}{S}_{b}^{ia^{\prime}}(-x)\,\big(\,(1-\gamma_{5})\gamma_{\mu}\big)\,\overset{\sim}{S}_{c}^{bi}(y)\gamma_{5}S_{u}^{ab^{\prime}}(y-x)]\nonumber\\
      		&&+\gamma_{5}\,S_{u}^{ca^{\prime}}(y-x)\,\gamma_{\nu}\,\overset{\sim}{S}_{d}^{bb^{\prime}}(y-x)\,S_{c}^{ai}(y)\,\big(\gamma_{\mu}\,(1-\gamma_{5})\big)\,S_{b}^{ic^{\prime}}(-x)\nonumber\\
      		&&-Tr[\,S_{d}^{ba^{\prime}}(y-x)\,\gamma_{\nu}\,\overset{\sim}{S}_{b}^{ib^{\prime}}(-x)\,\big(\,(1-\gamma_{5})\gamma_{\mu}\,\big)\overset{\sim}{S}_{c}^{ai}(y)]\,\gamma_{5}S_{u}^{cc^{\prime}}(y-x)\nonumber\\
      		&&-\gamma_{5}\,S_{u}^{cb^{\prime}}(y-x)\,\gamma_{\nu}\,\overset{\sim}{S}_{b}^{ia^{\prime}}(-x)\,\big(\,(1-\gamma_{5})\gamma_{\mu}\big)\,\overset{\sim}{S}_{c}^{ai}(y)S_{d}^{bc^{\prime}}(y-x)\nonumber\\
      		&&+\beta\,S_{u}^{ca^{\prime}}(y-x)\,\gamma_{\nu}\,\overset{\sim}{S}_{d}^{bb^{\prime}}(y-x)\gamma_{5}S_{c}^{ai}(y)\,\big(\gamma_{\mu}\,(1-\gamma_{5})\big)\,S_{b}^{ic^{\prime}}(-x)\nonumber\\
      		&&-Tr[\,S_{d}^{ba^{\prime}}(y-x)\,\gamma_{\nu}\,\overset{\sim}{S}_{b}^{ib^{\prime}}(-x)\,\big(\,(1-\gamma_{5})\gamma_{\mu}\,\big)\overset{\sim}{S}_{c}^{ai}(y)\gamma_{5}]\,S_{u}^{cc^{\prime}}(y-x)\nonumber\\
      		&&-\beta\,S_{u}^{cb^{\prime}}(y-x)\,\gamma_{\nu}\,\overset{\sim}{S}_{b}^{ia^{\prime}}(-x)\,\big(\,(1-\gamma_{5})\gamma_{\mu}\,\big)\overset{\sim}{S}_{c}^{ai}(y)\,\gamma_{5}S_{d}^{bc^{\prime}}(y-x)
      		\,\Bigg\rbrace,
      \end{eqnarray}
   where $\overset{\sim}{S}_{q}\,=\,C\,S^{T}\,C$. The light quark propagator is defined as \cite{Agaev:2020zad}
     \begin{eqnarray}
   	&&S_{q}^{ab}(x)\,=
    		i\,\delta_{ab}\frac{\slashed{x}}{2\pi^{2}x^{4}}\,-\,\delta_{ab}\frac{m_{q}}{4\pi^{2}x^{2}}\,-\,\delta_{ab}\,\frac{\langle\bar{q}q\rangle}{12}\,
    		+i\delta_{ab}\,\frac{\slashed{x}\,m_{q}\langle\bar{q}q\rangle}{48}\,-\,\delta_{ab}\,\frac{x^{2}}{192}\,\langle\bar{q}g_{s}\sigma G q\rangle\,\nonumber\\&&+i\delta_{ab}\,\frac{x^{2}\slashed{x}m_{q}}{1152}\,<\bar{q}g_{s}\sigma G q>\,
    		-i\frac{g_{s}\,G_{ab}^{\mu\nu}}{32\pi^{2}x^{2}}\,[\slashed{x}\,\sigma_{\mu\nu}+\sigma_{\mu\nu}\slashed{x}]\,-\,i\delta_{ab}\frac{x^{2}\slashed{x}g_{s}^{2}\langle\bar{q}q\rangle^{2}}{7776}\,-\,\delta_{ab}\frac{x^{4}\langle\bar{q}q\rangle\,\langle g_{s}^{2}G^{2}\rangle}{27648}\,+\,\cdots ,
    \end{eqnarray}     
    and the heavy quark propagator has the following representation \cite{Agaev:2020zad}:    
      \begin{eqnarray}
     		&&S_{Q}^{ab}(x)\,=\,
     		i\int\,\frac{\mathrm{d}^{4}k}{(2\pi)^{4}}\,e^{-ikx}\,\Bigg\lbrace\,\frac{\delta_{ab}(\slashed{k}+m_{Q})}{k^{2}-m_{Q}^{2}}-\,\frac{g_{s}\,G_{ab}^{\mu\nu}}{4}\,\frac{\sigma_{\mu\nu}(\slashed{k}+m_{Q})\,+\,(\slashed{k}+m_{Q})\sigma_{\mu\nu}}{(k^{2}-m_{Q}^{2})^{2}}\nonumber\\
     		&&\,+\,\frac{g_{s}^{2}G^{2}}{12}\delta_{ab}\,m_{Q}\frac{k^{2}+m_{Q}\slashed{k}}{(k^{2}-m_{Q}^{2})^{4}}\,+\,\frac{g_{s}^{3}G^{3}}{48}\delta_{ab}\,\frac{(\slashed{k}+m_{Q})}{(k^{2}-m_{Q}^{2})^{6}}[\,\slashed{k}(k^{2}-3m_{Q}^{2})\,+\,2m_{Q}(2k^{2}-m_{Q}^{2})](\slashed{k}+m_{Q})+\cdots\,\Bigg\rbrace\, ,
     \end{eqnarray}
     where $m_{q}$ and $m_{Q}$ indicate the light and heavy quark masses and $k$ represents the four-momentum of heavy quark. We also define
     \begin{equation}
     	G_{ab}^{\mu\nu}\,\equiv\,G_{A}^{\mu\nu}\,t_{ab}^{A},\,\,\,\,\,G^{2}=G_{\mu\nu}^{A}\,G_{A}^{\mu\nu},
     \end{equation}
     with , $t^{A}=\lambda^{A}/2$ where $\lambda^{A}$ denotes the Gell-Mann matrices with $A=1,2...8$ and $\mu$ and $\nu$ are Lorentz indices. The gluon field strength tensor $G_{\mu\nu}^{A}=G_{\mu\nu}^{A}(0)$ is fixed at $x=0$. 
     
     By substituting the heavy and light quark propagators in Eq. \eqref{qcdside}, we obtain all  the perturbative and nonperturbative contributions with different mass dimensions in the OPE. The lowest-dimension operator, $d=0$, is related to the perturbative contribution and the high-dimensional operators correspond to the non-perturbative effects. They are associated with the vacuum condensates, such as quark condensate with $d=3,\,\langle \bar{q}q \rangle$, gluon condensate, $d=4,\, \langle G^{2}\rangle $, quark-gluon condensate, $d=5,\, \langle \bar{q}g\sigma G q\rangle $, and two-times quark condensate, $d=6,\, \langle \bar{q}q \rangle^{2}$. In this study, the nonperturbative effects up to six mass dimensions are considered in the calculations. 
     
     Our study proceeds by evaluating the four-integrals of the form,
     \begin{equation}
     \int \mathrm{d}^{4}k\,\int \mathrm{d}^{4}k^{\prime}\,\int \mathrm{d}^{4}x\,e^{i\,(k-p).x}\,\int \mathrm{d}^{4}y\,e^{i\,(-k^{\prime}+p^{\prime}).y}\,\frac{x_{\mu}\,y_{\nu}\,k_{\alpha}\,k^{\prime}_{\beta}}{(k^{2}-m_{b}^{2})^{l}\,(k^{\prime 2}-m_{c}^{2})^{m}\,[(y-x)^{2}]^{n}}.
     \end{equation} 
    In the first step, we utilize the identity \cite{Azizi:2017ubq},
    \begin{equation}
    \frac{1}{[(y-x)^{2}]^{n}}\,=\,\int \, \frac{\mathrm{d}^{D}t}{(2\pi)^{D}}\,e^{-i\,t.(y-x)}\,i\,(-1)^{n+1}\,2^{D-2n}\,\pi^{D/2}\,\frac{\Gamma(D/2-n)}{\Gamma(n)}\,(-\frac{1}{t^{2}})^{D/2-n},
    \end{equation}      
    to write a part of the denominators in exponential forms.
    
    After substituting $x_{\mu}\rightarrow i\,\frac{\partial}{\partial p_{\mu}}$ and $y_{\mu}\rightarrow -i\,\frac{\partial}{\partial p^{\prime}_{\mu}}$ and performing the Fourier integrals by employing the definition of the four-dimensional Dirac delta function,  we evaluate the integrals over four $k$ and $k^{\prime}$. The ultimate expressions are then computed via the Feynman parametrization and the following formula \cite{Azizi:2017ubq}:
    \begin{equation}
    \int \mathrm{d}^{D}t\,\frac{(t^2)^m}{(t^2+\Delta)^n}\,=\,\frac{i\,\pi^{2}\,(-1)^{m-n}\,\Gamma(m+2)\,\Gamma(n-m-2)}{\Gamma(2)\,\Gamma(n)\,[-\Delta]^{n-m-2}},
    \end{equation}
    and the imaginary parts of the results are extracted by applying the identity \cite{Azizi:2017ubq},
     \begin{equation}
     \Gamma[\frac{D}{2}-n]\,(-\frac{1}{\Delta})^{D/2-n}\,=\,\frac{(-1)^{n-1}}{(n-2)!}\,(-\Delta)^{n-2}\,ln[-\Delta].	
     \end{equation} 
    Finally, the QCD side of the correlation function is acquired in terms of thirty-two different Lorentz structures as follows:
    \begin{eqnarray}\label{QCD1}
    &&\Pi_{\mu\nu}^{QCD}(p,p^{\prime},q^{2})\,=\,\Pi_{p_{\mu}p^{\prime}_{\nu}}^{QCD}(p^{2},p^{\prime2},q^{2})\,p_{\mu}p^{\prime}_{\nu}\,+\,\Pi_{p^{\prime}_{\mu}p^{\prime}_{\nu}}^{QCD}(p^{2},p^{\prime2},q^{2})\,p^{\prime}_{\mu}p^{\prime}_{\nu}\,+\,\Pi_{g_{\mu\nu}}^{QCD}(p^{2},p^{\prime2},q^{2})\,g_{\mu\nu}\,+\,\Pi_{p_{\mu}p^{\prime}_{\nu}\gamma_{5}}^{QCD}(p^{2},p^{\prime2},q^{2}) \,\notag\\ &&p_{\mu}p^{\prime}_{\nu}\gamma_{5}\,+\,\Pi_{p^{\prime}_{\mu}p^{\prime}_{\nu}\gamma_{5}}^{QCD}(p^{2},p^{\prime2},q^{2})\,p^{\prime}_{\mu}p^{\prime}_{\nu}\gamma_{5}\,+\,\Pi_{g_{\mu\nu}\gamma_{5}}^{QCD}(p^{2},p^{\prime2},q^{2})\,g_{\mu\nu}\gamma_{5}\,+\,\Pi_{p_{\mu}p^{\prime}_{\nu}\slashed{p}}^{QCD}(p^{2},p^{\prime2},q^{2}) \,p_{\mu}p^{\prime}_{\nu}\slashed{p}\,+\,\Pi_{p^{\prime}_{\mu}p^{\prime}_{\nu}\slashed{p}}^{QCD}(p^{2},p^{\prime2},q^{2})\notag\\&& \,p^{\prime}_{\mu}p^{\prime}_{\nu}\slashed{p}\,+\,\Pi_{g_{\mu\nu}\slashed{p}}^{QCD}(p^{2},p^{\prime2},q^{2}) \,g_{\mu\nu}\slashed{p}\,+\,\Pi_{p_{\mu}p^{\prime}_{\nu}\slashed{p}^{\prime}}^{QCD}(p^{2},p^{\prime2},q^{2}) \,p_{\mu}p^{\prime}_{\nu}\slashed{p}^{\prime}\,+\,\Pi_{p^{\prime}_{\mu}p^{\prime}_{\nu}\slashed{p}^{\prime}}^{QCD}(p^{2},p^{\prime2},q^{2}) \,p^{\prime}_{\mu}p^{\prime}_{\nu}\slashed{p}^{\prime}\,+\,\Pi_{g_{\mu\nu}\slashed{p}^{\prime}}^{QCD}(p^{2},p^{\prime2},q^{2})\notag\\&& \,g_{\mu\nu}\slashed{p}^{\prime}\,+\,\Pi_{p^{\prime}_{\nu}\gamma_{\mu}}^{QCD}(p^{2},p^{\prime2},q^{2}) \,p^{\prime}_{\nu}\gamma_{\mu}\,+\,\Pi_{p_{\mu}p^{\prime}_{\nu}\slashed{p}\gamma_{5}}^{QCD}(p^{2},p^{\prime2},q^{2}) \,p_{\mu}p^{\prime}_{\nu}\slashed{p}\gamma_{5}\,+\,\Pi_{p^{\prime}_{\mu}p^{\prime}_{\nu}\slashed{p}\gamma_{5}}^{QCD}(p^{2},p^{\prime2},q^{2}) \,p^{\prime}_{\mu}p^{\prime}_{\nu}\slashed{p}\gamma_{5}\,+\,\Pi_{g_{\mu\nu}\slashed{p}\gamma_{5}}^{QCD}(p^{2},p^{\prime2},q^{2})\notag\\&& \,g_{\mu\nu}\slashed{p}\gamma_{5}\,+\,\Pi_{p_{\mu}p^{\prime}_{\nu}\slashed{p}\slashed{p}^{\prime}}^{QCD}(p^{2},p^{\prime2},q^{2}) \,p_{\mu}p^{\prime}_{\nu}\slashed{p}\slashed{p}^{\prime}\,+\,\Pi_{p^{\prime}_{\mu}p^{\prime}_{\nu}\slashed{p}\slashed{p}^{\prime}}^{QCD}(p^{2},p^{\prime2},q^{2}) \,p^{\prime}_{\mu}p^{\prime}_{\nu}\slashed{p}\slashed{p}^{\prime}\,+\,\Pi_{g_{\mu\nu}\slashed{p}\slashed{p}^{\prime}}^{QCD}(p^{2},p^{\prime2},q^{2}) \,g_{\mu\nu}\slashed{p}\slashed{p}^{\prime}\,+\,\notag\\&&\Pi_{p_{\mu}p^{\prime}_{\nu}\slashed{p}^{\prime}\gamma_{5}}^{QCD}(p^{2},p^{\prime2},q^{2}) \,p_{\mu}p^{\prime}_{\nu}\slashed{p}^{\prime}\gamma_{5}\,+\,\Pi_{p^{\prime}_{\mu}p^{\prime}_{\nu}\slashed{p}^{\prime}\gamma_{5}}^{QCD}(p^{2},p^{\prime2},q^{2}) \,p^{\prime}_{\mu}p^{\prime}_{\nu}\slashed{p}^{\prime}\gamma_{5}\,+\,\Pi_{g_{\mu\nu}\slashed{p}^{\prime}\gamma_{5}}^{QCD}(p^{2},p^{\prime2},q^{2}) \,g_{\mu\nu}\slashed{p}^{\prime}\gamma_{5}\,+\,\notag\\&&\Pi_{p^{\prime}_{\nu}\gamma_{\mu}\gamma_{5}}^{QCD}(p^{2},p^{\prime2},q^{2}) \,p^{\prime}_{\nu}\gamma_{\mu}\gamma_{5}\,+\,\Pi_{p^{\prime}_{\nu}\gamma_{\mu}\slashed{p}}^{QCD}(p^{2},p^{\prime2},q^{2}) \,p^{\prime}_{\nu}\gamma_{\mu}\slashed{p}\,+\,\Pi_{p^{\prime}_{\nu}\gamma_{\mu}\slashed{p}^{\prime}}^{QCD}(p^{2},p^{\prime2},q^{2}) \,p^{\prime}_{\nu}\gamma_{\mu}\slashed{p}^{\prime}\,+\,\notag\\&&\Pi_{p_{\mu}p^{\prime}_{\nu}\slashed{p}\slashed{p}^{\prime}\gamma_{5}}^{QCD}(p^{2},p^{\prime2},q^{2}) \,p_{\mu}p^{\prime}_{\nu}\slashed{p}\slashed{p}^{\prime}\gamma_{5}\, +\Pi_{p^{\prime}_{\mu}p^{\prime}_{\nu}\slashed{p}\slashed{p}^{\prime}\gamma_{5}}^{QCD}(p^{2},p^{\prime2},q^{2}) \,p^{\prime}_{\mu}p^{\prime}_{\nu}\slashed{p}\slashed{p}^{\prime}\gamma_{5}\,+\,\Pi_{g_{\mu\nu}\slashed{p}\slashed{p}^{\prime}\gamma_{5}}^{QCD}(p^{2},p^{\prime2},q^{2}) \,g_{\mu\nu}\slashed{p}\slashed{p}^{\prime}\gamma_{5}\,+\,\notag\\&&\Pi_{p^{\prime}_{\nu}\gamma_{\mu}\slashed{p}\gamma_{5}}^{QCD}(p^{2},p^{\prime2},q^{2}) \,p^{\prime}_{\nu}\gamma_{\mu}\slashed{p}\gamma_{5}\,+\,\Pi_{p^{\prime}_{\nu}\gamma_{\mu}\slashed{p}\slashed{p}^{\prime}}^{QCD}(p^{2},p^{\prime2},q^{2}) \,p^{\prime}_{\nu}\gamma_{\mu}\slashed{p}\slashed{p}^{\prime}\,+\,\Pi_{p^{\prime}_{\nu}\gamma_{\mu}\slashed{p}^{\prime}\gamma_{5}}^{QCD}(p^{2},p^{\prime2},q^{2}) \,p^{\prime}_{\nu}\gamma_{\mu}\slashed{p}^{\prime}\gamma_{5}\,+\,\notag\\&&\Pi_{p^{\prime}_{\nu}\gamma_{\mu}\slashed{p}\slashed{p}^{\prime}\gamma_{5}}^{QCD}(p^{2},p^{\prime2},q^{2}) \,p^{\prime}_{\nu}\gamma_{\mu}\slashed{p}\slashed{p}^{\prime}\gamma_{5}\,,
    \end{eqnarray}
    where the invariant functions $\Pi_{i}^{QCD}(p^{2},p^{\prime2},q^{2})$, ($i$ denotes individual structures) are described in terms of double dispersion integrals,
    \begin{equation}
    \Pi_{i}^{QCD}(p^{2},p^{\prime2},q^{2})\,=\,\int_{s_{min}}^{\infty}\,\mathrm{d}s\,\int_{s^{\prime}_{min}}^{\infty}\,\mathrm{d}s^{\prime}\,\frac{\rho_{i}^{QCD}(s,s^{\prime},q^{2})}{(s-p^{2})(s^{\prime}-p^{\prime2})}	\,+\,\Gamma_{i}(p^{2},p^{\prime2},q^{2}),
    \end{equation}
    where $s_{min}=m_{b}^{2}$, $s^{\prime}_{min}=m_{c}^{2}$, and $\rho_{i}^{QCD}(s,s^{\prime},q^{2})$ indicates the spectral densities, obtained by $\rho_{i}^{QCD}(s,s^{\prime},q^{2})\,=\,\frac{1}{\pi}\,Im\,\Pi_{i}^{QCD}(p^{2},p^{\prime2},q^{2})$. Here, $\Gamma_{i}(p^{2},p^{\prime2},q^{2})$ denotes the contributions with no imaginary parts that are directly computed. By employing the quark-hadron duality assumption, the integrals' upper limits will be turned to $s_{0}$ and $s^{\prime}_{0}$ that are known as continuum thresholds of the initial and final baryons, respectively. In calculations, the spectral densities decompose as follows:
    \begin{equation}
    \rho_{i}^{QCD}(s,s^{\prime},q^{2})\,=\,\rho_{i}^{Pert.}(s,s^{\prime},q^{2})\,+\,\sum_{n=3}^{6}\,\rho_{i}^{n}(s,s^{\prime},q^{2}),	 
    \end{equation} 
    where the first part, $\rho_{i}^{Pert.}(s,s^{\prime},q^{2})$, is related to the perturbative contributions and the second part, as mentioned above, denotes the nonperturbative effects and vacuum condensates including quark, gluon, and mixed condensates. In this step, the contributions of the higher states and continuum should be suppressed by applying the double Borel transformation, Eq.\eqref{Borel}, to the QCD side and employing continuum subtraction arisen from the quark-hadron duality assumption. Therefore, we acquire
    \begin{equation}
    \Pi_{i}^{QCD}\,(M^{2},M^{\prime2},s_{0},s^{\prime}_{0},q^{2})\,=\,\int_{s_{min}}^{s_{0}}\,\mathrm{d}s\,\int_{s^{\prime}_{min}}^{s^{\prime}_{0}}\,\mathrm{d}s^{\prime}\,e^{-s/M^{2}}\,e^{-s^{\prime}/M^{\prime2}}\,\rho_{i}^{QCD}(s,s^{\prime},q^{2})	\,+\,\mathbf{\widehat{B}}\big[\Gamma_{i}(p^{2},p^{\prime2},q^{2})\big],
    \end{equation}   
    where the final expressions for the spectral densities, $\rho_{i}(s,s^{\prime},q^{2})$, and functions $\Gamma_{i}(p^{2},p^{\prime2},q^{2})$ corresponding to the structure $g_{\mu\nu}\,\slashed{p}$ are provided in Appendix \ref{appB}.
    
    Ultimately, by matching the corresponding coefficients of the distinctive Lorentz structures from the physical and theoretical sides, we acquire the QCD‌ sum rules to determine the responsible form factors, in terms of QCD fundamental parameters, such as the strong coupling constant; quark,  gluon, and mixed condensates; quark masses as well as hadronic parameters, such as residues and masses  of the initial and final baryons; and  finally, auxiliary parameters, $s_{0},\, s^{\prime}_{0},\, M^{2},\, M^{\prime2}$, and $\beta$.  The expressions of the sum rules for the form factors are presented in Appendix \ref{app new}.     
    \section{NUMERICAL ANALYSIS OF THE FORM FACTORS} \label{analysis}
    As mentioned in Sec. \ref{sec1}, the study of the heavy baryons and their decay properties is of crucial importance in probing hadronic structures. The decay width, as an observable quantity in experiment, is a main source to achieve this purpose. In theory, the decay width is obtained by computing the responsible form factors. In our approach, form factors are determined after acquiring the QCD sum rules and analyzing the numerical results. Since the form factors are invariant functions of $q^{2}$, it is essential to analyze their behavior in terms of $q^{2}$. To this end, some input parameters are required that are itemized in Table \ref{inputpara}.
    
    \begin{table}[h]
    \centering
     \caption{The input parameters used in numerical analysis.}
    \begin{tabular}{|c|c|}
    	\hline
    	Parameters&Values\\
    	\hline
    	$m_{b}$&  $\,(\,4.18^{+0.03}_{-0.02}\,)\,\,\mathrm{GeV}$ \cite{ParticleDataGroup:2024cfk}\\
    	$m_{c}$&  $\,(1.27\,\pm\,0.02)\,\,\mathrm{GeV}$\cite{ParticleDataGroup:2024cfk}\\
    	$m_{e}$&  $0.51 \,\,\, \mathrm{MeV}$ \cite{ParticleDataGroup:2024cfk}\\
    	$m_{\mu}$&  $105.65 \,\,\, \mathrm{MeV}$ \cite{ParticleDataGroup:2024cfk}\\
    	$m_{\tau}$&  $1776.93 \,\,\, \mathrm{MeV}$ \cite{ParticleDataGroup:2024cfk}\\
    	$m_{\Sigma_{b}^{*}}$&    $\,\,\,(5830.32\,\pm\,0.27)\,\,\,  \mathrm{MeV}$ \cite{ParticleDataGroup:2024cfk}\\
    	$m_{\Sigma_{c}^{+}}$&$\,\,\,(2452.65^{+0.22}_{-0.16})\,\,\,   \mathrm{MeV}$ \cite{ParticleDataGroup:2024cfk}\\
    	$\lambda_{\Sigma_{b}^{*}}$&$\,\,(0.038\,\pm\,0.011)\,\, \mathrm{GeV^{3}}$ \cite{Wang:2010vn}\\
    	$\lambda_{\Sigma_{c}^{+}}$&$\,\,(0.045\,\pm\,0.015)\,\, \mathrm{GeV^{3}}$ \cite{Wang:2009cr}\\
    	$G_{F}$&$\,\,1.17\times 10^{-5}\,\,\,\mathrm{GeV^{-2}}\,\,$ \cite{ParticleDataGroup:2024cfk}\\
    	$V_{cb}$&$\,\,(39\pm 1.1)\times 10^{-3}\,\,$ \cite{ParticleDataGroup:2024cfk}\\
    	$m_{0}^{2}$&$\,\,(0.8 \pm 0.2)\,\,\,\mathrm{GeV^{2}}\,\,$ \cite{Ioffe:2005ym, Belyaev:1982cd, Belyaev:1982sa}\\
    	$\langle \bar{u}u \rangle$&$\,\, -(0.24\pm 0.01)^{3}\,\,\mathrm{GeV^{3}}$ \cite{Belyaev:1982cd, Belyaev:1982sa}\\
    	$\,\,\langle 0\,|\,\frac{1}{\pi}\alpha_{s}\mathrm{G^{2}}\,|\,0 \rangle\,\,$& $\,\,\,(0.012\pm 0.004)\,\,\mathrm{GeV^{4}}\,\,\,$ \cite{Ioffe:2005ym, Belyaev:1982cd, Belyaev:1982sa}\\
    	\hline
    \end{tabular}
    \label{inputpara}
    \end{table}
    
\begin{figure}
	 	\begin{center}
	 	\subfigure{\includegraphics[height=5 cm , width=5.5 cm]{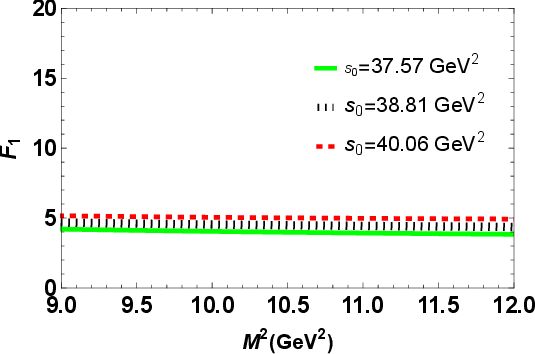}}
	 	\hspace*{0.02 cm}	
	 	\subfigure{\includegraphics[height=5 cm , width=5.5 cm]{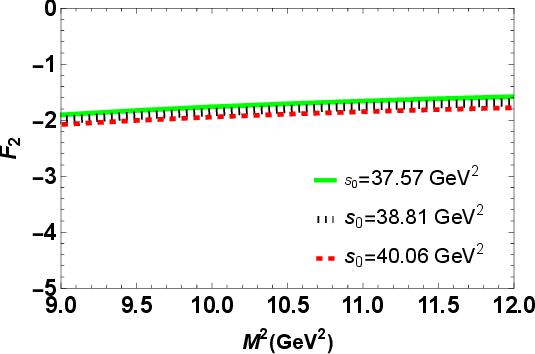}}
	 	\hspace*{0.02 cm}
	 	\subfigure{\includegraphics[height=5 cm , width=5.5 cm]{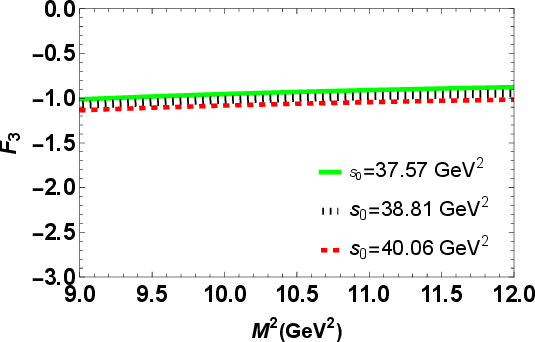}}
	 	
	 	\subfigure{\includegraphics[height=5 cm , width=5.5 cm]{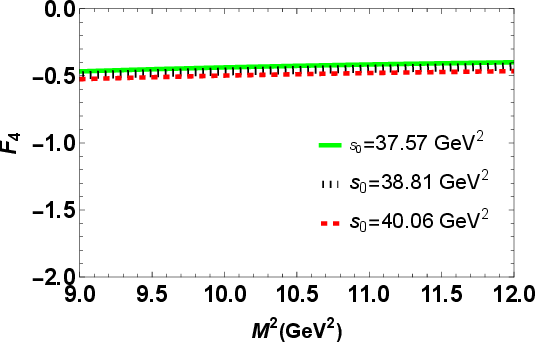}}
	 	\hspace*{0.02 cm}
	 	\subfigure{\includegraphics[height=5 cm , width=5.5 cm]{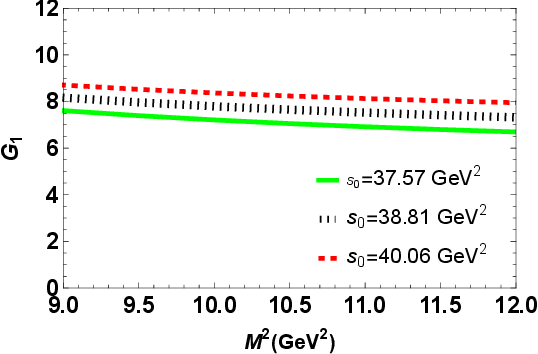}}
	 	\hspace*{0.02 cm}
	 	\subfigure{\includegraphics[height=5 cm , width=5.5 cm]{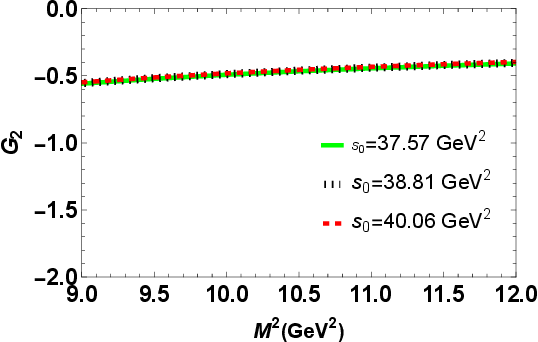}}
	 	
	 	\subfigure{\includegraphics[height=5 cm , width=5.5 cm]{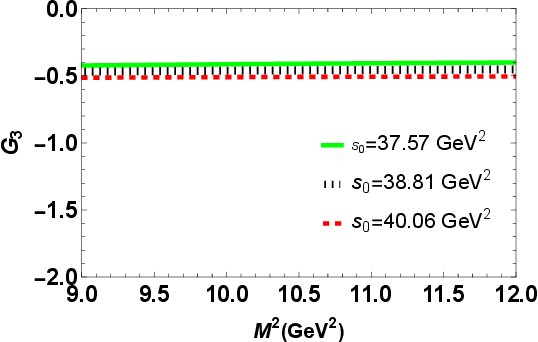}}
	 	\hspace*{0.02 cm}
	 	\subfigure{\includegraphics[height=5 cm , width=5.5 cm]{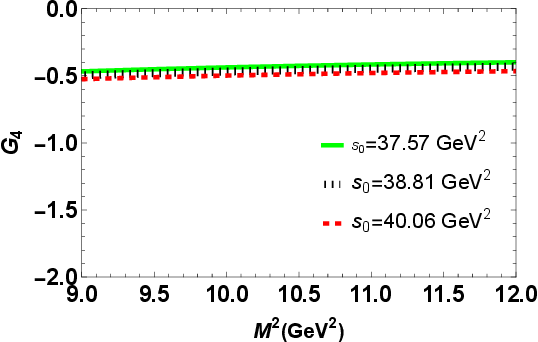}}
	 	\end{center}
	 	\caption{Variations of the form factors with  respect to  the Borel parameter $M^{2}$ for a  variety of the parameter $s_{0}$, $q^{2}=0$, and the central values of the other auxiliary parameters at Ioffe point. The figures relate to the structures $g_{\mu\nu}\slashed{p}\gamma_{5},\,p^{\prime}_{\nu}\gamma_{\mu}\slashed{p}\slashed{p}^{\prime}\gamma_{5},\,p_{\mu}p^{\prime}_{\nu}\slashed{p}\slashed{p}^{\prime}\gamma_{5},\,p^{\prime}_{\mu}p^{\prime}_{\nu}\slashed{p}\slashed{p}^{\prime}\gamma_{5},\,g_{\mu\nu}\slashed{p},\,p^{\prime}_{\mu}p^{\prime}_{\nu}\slashed{p},\,p_{\mu}p^{\prime}_{\nu}\slashed{p}^{\prime}$, and $p^{\prime}_{\mu}p^{\prime}_{\nu}\slashed{p}\slashed{p}^{\prime} $ corresponding to the form factors $F_{1},\, F_{2},\, F_{3},\, F_{4},\, G_{1},\, G_{2},\, G_{3}$, and $G_{4}$, respectively (see Table \ref{set1}).} 
	 	\label{ffm2s0}
	 \end{figure}

\begin{figure}
 		\begin{center}
 			\subfigure{\includegraphics[height=5 cm , width=5.5 cm]{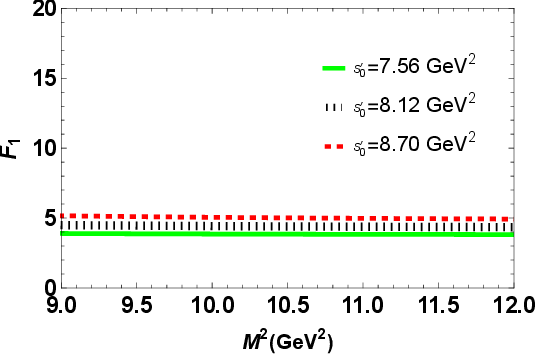}}
 			\hspace*{0.02 cm}	
 			\subfigure{\includegraphics[height=5 cm , width=5.5 cm]{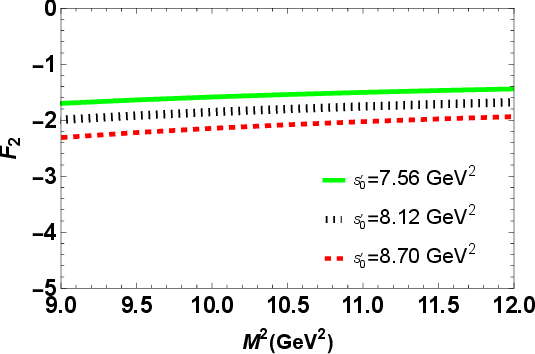}}
 			\hspace*{0.02 cm}
 			\subfigure{\includegraphics[height=5 cm , width=5.5 cm]{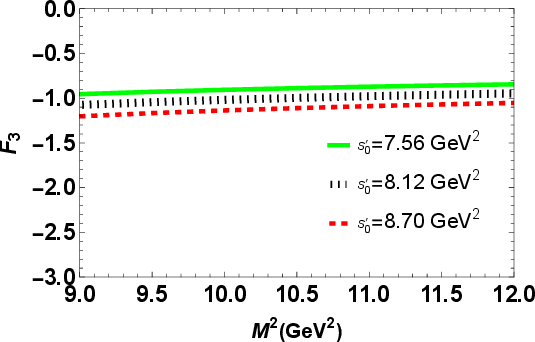}}
 			
 			\subfigure{\includegraphics[height=5 cm , width=5.5 cm]{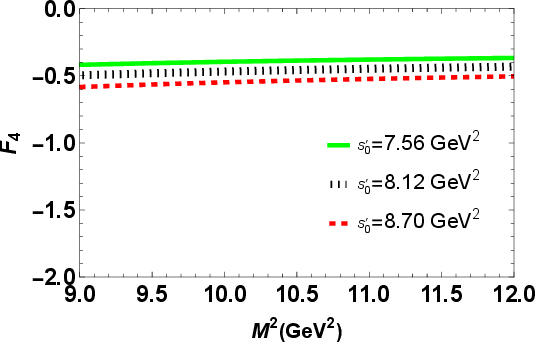}}
 			\hspace*{0.02 cm}
 			\subfigure{\includegraphics[height=5 cm , width=5.5 cm]{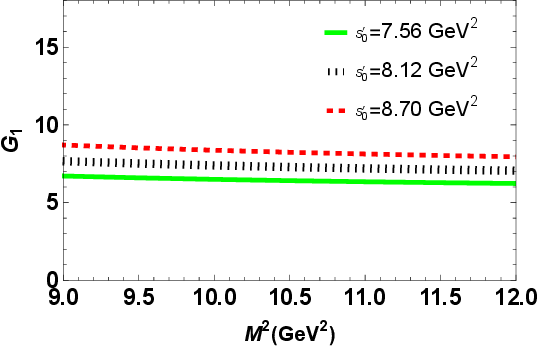}}
 			\hspace*{0.02 cm}
 			\subfigure{\includegraphics[height=5 cm , width=5.5 cm]{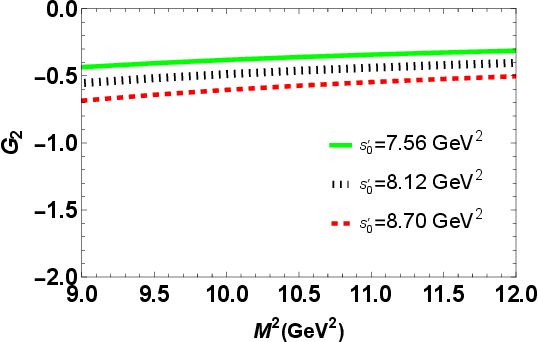}}
 			
 			\subfigure{\includegraphics[height=5 cm , width=5.5 cm]{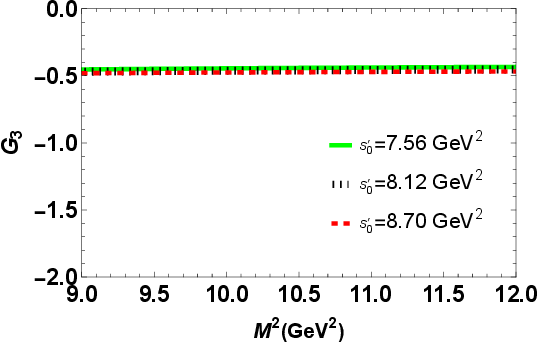}}
 			\hspace*{0.02 cm}
 			\subfigure{\includegraphics[height=5 cm , width=5.5 cm]{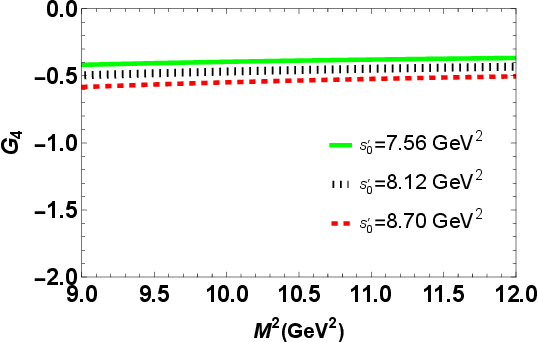}}
 		\end{center}
 		\caption{Variations of the form factors with respect to  the Borel parameter $M^{2}$ for a  variety of the parameter $s_{0}^{\prime}$, $q^{2}=0$, and the central values of the other auxiliary parameters at Ioffe point. The figures relate to the structures $g_{\mu\nu}\slashed{p}\gamma_{5},\,p^{\prime}_{\nu}\gamma_{\mu}\slashed{p}\slashed{p}^{\prime}\gamma_{5},\,p_{\mu}p^{\prime}_{\nu}\slashed{p}\slashed{p}^{\prime}\gamma_{5},\,p^{\prime}_{\mu}p^{\prime}_{\nu}\slashed{p}\slashed{p}^{\prime}\gamma_{5},\,g_{\mu\nu}\slashed{p},\,p^{\prime}_{\mu}p^{\prime}_{\nu}\slashed{p},\,p_{\mu}p^{\prime}_{\nu}\slashed{p}^{\prime}$, and $p^{\prime}_{\mu}p^{\prime}_{\nu}\slashed{p}\slashed{p}^{\prime} $ corresponding to the form factors $F_{1},\, F_{2},\, F_{3},\, F_{4},\, G_{1},\, G_{2},\, G_{3}$, and $G_{4}$, respectively (see Table \ref{set1}).}
 		\label{ffm2sp0}
 	\end{figure}

\begin{figure}
	 	\begin{center}
	 		\subfigure{\includegraphics[height=5 cm , width=5.5 cm]{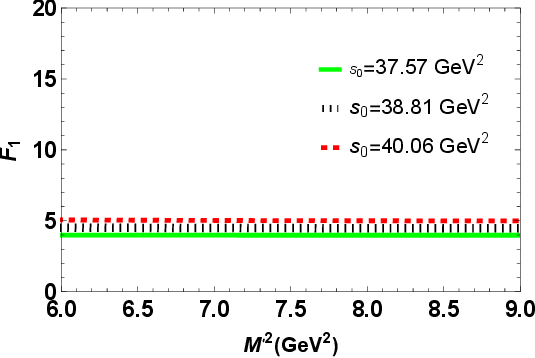}}
	 		\hspace*{0.02 cm}	
	 		\subfigure{\includegraphics[height=5 cm , width=5.5 cm]{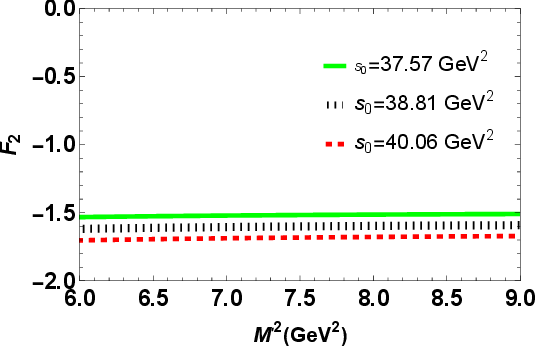}}
	 		\hspace*{0.02 cm}
	 		\subfigure{\includegraphics[height=5 cm , width=5.5 cm]{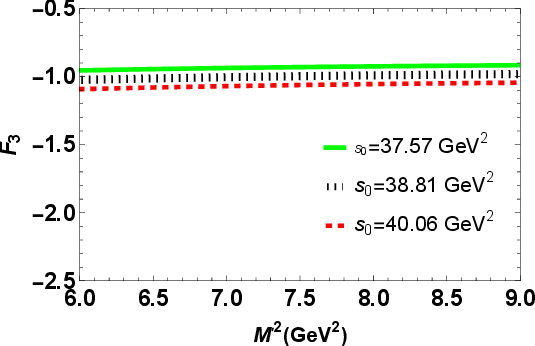}}
	 		
	 		\subfigure{\includegraphics[height=5 cm , width=5.5 cm]{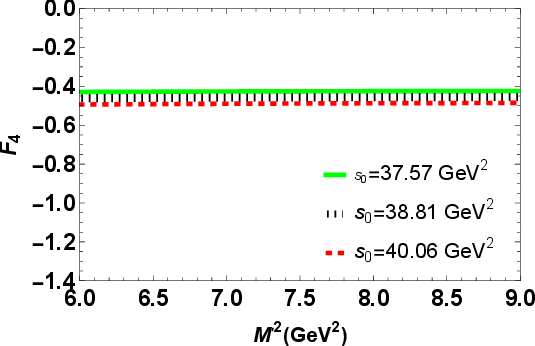}}
	 		\hspace*{0.02 cm}
	 		\subfigure{\includegraphics[height=5 cm , width=5.5 cm]{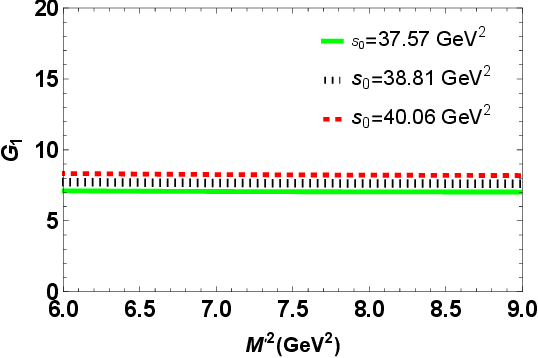}}
	 		\hspace*{0.02 cm}
	 		\subfigure{\includegraphics[height=5 cm , width=5.5 cm]{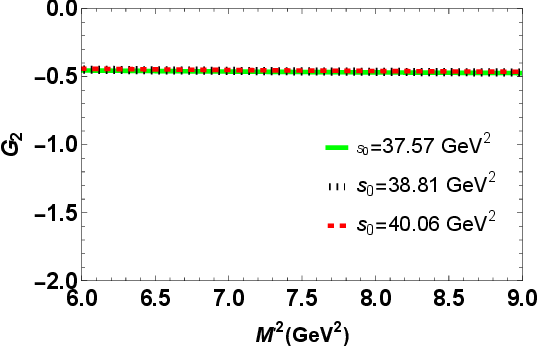}}
	 		
	 		\subfigure{\includegraphics[height=5 cm , width=5.5 cm]{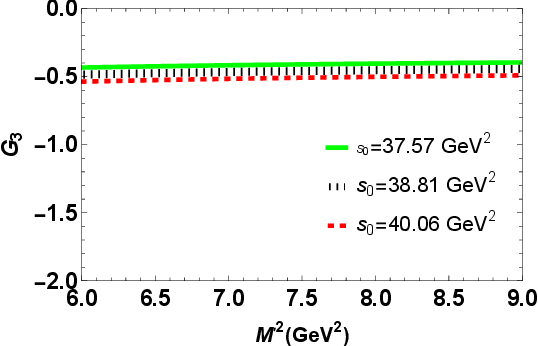}}
	 		\hspace*{0.02 cm}
	 		\subfigure{\includegraphics[height=5 cm , width=5.5 cm]{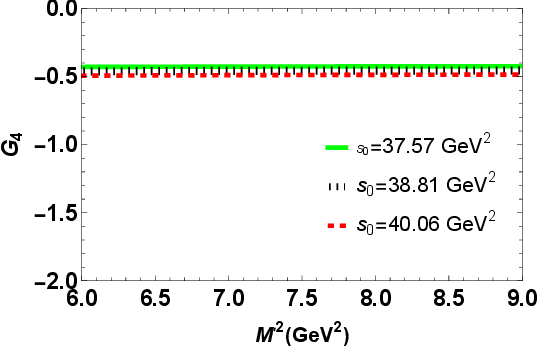}}
	 	\end{center}
	 	\caption{Variations of the form factors with respect to the Borel parameter $M^{\prime2}$ for a  variety of the parameter $s_{0}$, $q^{2}=0$, and the central values of the other auxiliary parameters at Ioffe point. The figures relate to the structures $g_{\mu\nu}\slashed{p}\gamma_{5},\,p^{\prime}_{\nu}\gamma_{\mu}\slashed{p}\slashed{p}^{\prime}\gamma_{5},\,p_{\mu}p^{\prime}_{\nu}\slashed{p}\slashed{p}^{\prime}\gamma_{5},\,p^{\prime}_{\mu}p^{\prime}_{\nu}\slashed{p}\slashed{p}^{\prime}\gamma_{5},\,g_{\mu\nu}\slashed{p},\,p^{\prime}_{\mu}p^{\prime}_{\nu}\slashed{p},\,p_{\mu}p^{\prime}_{\nu}\slashed{p}^{\prime}$, and $p^{\prime}_{\mu}p^{\prime}_{\nu}\slashed{p}\slashed{p}^{\prime} $ corresponding to the form factors $F_{1},\, F_{2},\, F_{3},\, F_{4},\, G_{1},\, G_{2},\, G_{3}$, and $G_{4}$, respectively (see Table \ref{set1}).}
	 	\label{ffmp2s0}
	 \end{figure}

\begin{figure}
	 	\begin{center}
	 		\subfigure{\includegraphics[height=5 cm , width=5.5 cm]{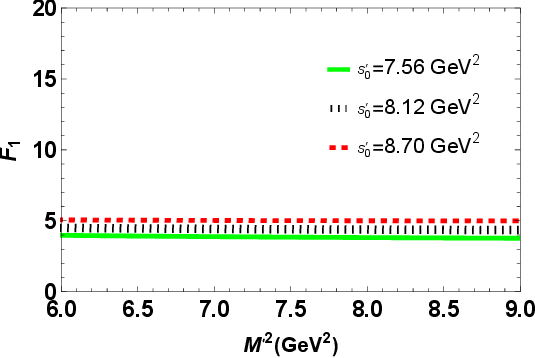}}
	 		\hspace*{0.02 cm}	
	 		\subfigure{\includegraphics[height=5 cm , width=5.5 cm]{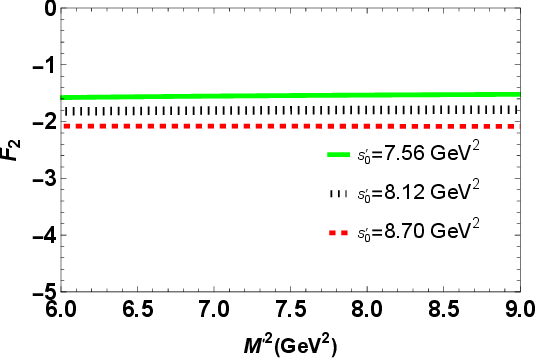}}
	 		\hspace*{0.02 cm}
	 		\subfigure{\includegraphics[height=5 cm , width=5.5 cm]{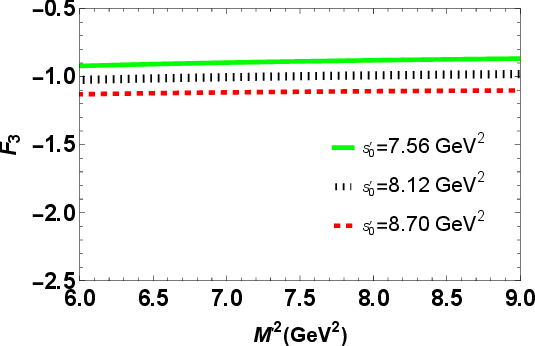}}
	 		
	 		\subfigure{\includegraphics[height=5 cm , width=5.5 cm]{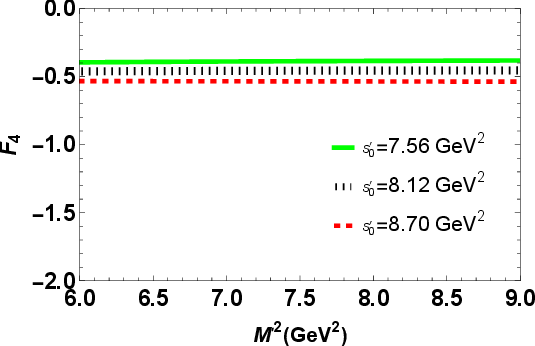}}
	 		\hspace*{0.02 cm}
	 		\subfigure{\includegraphics[height=5 cm , width=5.5 cm]{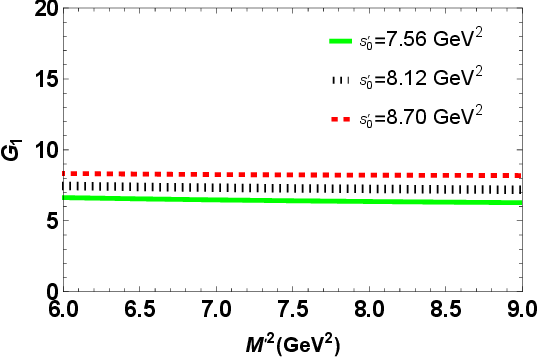}}
	 		\hspace*{0.02 cm}
	 		\subfigure{\includegraphics[height=5 cm , width=5.5 cm]{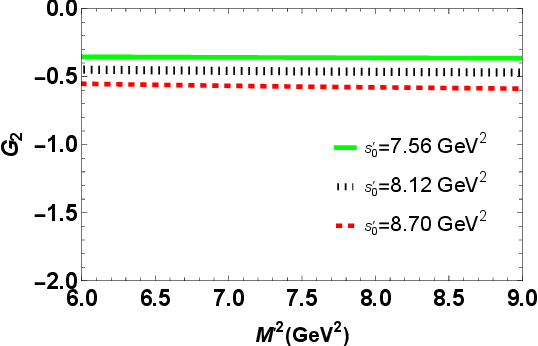}}
	 		
	 		\subfigure{\includegraphics[height=5 cm , width=5.5 cm]{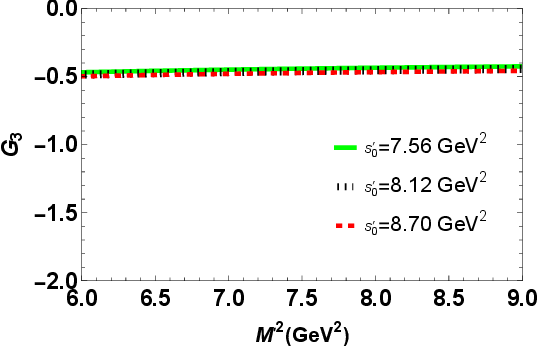}}
	 		\hspace*{0.02 cm}
	 		\subfigure{\includegraphics[height=5 cm , width=5.5 cm]{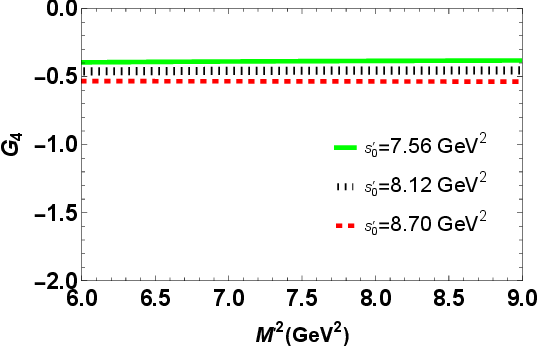}}
	 	\end{center}
	 	\caption{Variations of the form factors with  respect to  the Borel parameter $M^{\prime2}$ for a  variety of the parameter $s^{\prime}_{0}$, $q^{2}=0$, and the central values of the other auxiliary parameters at Ioffe point. The figures relate to the structures $g_{\mu\nu}\slashed{p}\gamma_{5},\,p^{\prime}_{\nu}\gamma_{\mu}\slashed{p}\slashed{p}^{\prime}\gamma_{5},\,p_{\mu}p^{\prime}_{\nu}\slashed{p}\slashed{p}^{\prime}\gamma_{5},\,p^{\prime}_{\mu}p^{\prime}_{\nu}\slashed{p}\slashed{p}^{\prime}\gamma_{5},\,g_{\mu\nu}\slashed{p},\,p^{\prime}_{\mu}p^{\prime}_{\nu}\slashed{p},\,p_{\mu}p^{\prime}_{\nu}\slashed{p}^{\prime}$, and $p^{\prime}_{\mu}p^{\prime}_{\nu}\slashed{p}\slashed{p}^{\prime} $ corresponding to the form factors $F_{1},\, F_{2},\, F_{3},\, F_{4},\, G_{1},\, G_{2},\, G_{3}$, and $G_{4}$, respectively ( see Table \ref{set1}).}
	 	\label{ffmp2sp0}
	 \end{figure}

     In addition, as we previously mentioned, some auxiliary parameters appear in the calculations. These are the Borel parameters $M^{2}$ and $M^{\prime2}$, the continuum thresholds $s_{0}$ and $s^{\prime}_{0}$, and the mixing parameter $\beta$ in the interpolating currents of spin-$\frac{1}{2}$ baryons. Based on the standard requirements of the method, physical quantities are supposed to possibly show  stable behavior with respect to variations of these helping parameters. This constraint in addition to other  requirements, including pole dominance, at the initial and final channels, and convergence of the OPE, impose restrictions on the working domains of these parameters. In this framework, the contribution of the ground state, i.e. pole contribution, is considered to be larger than the contributions of the higher states and continuum. This consideration introduces the upper limits of the Borel parameters, $M^{2}$ and $M^{\prime2}$ by demanding 
         \begin{equation}
         	PC=\frac{\Pi^{QCD}(M^{2},M^{\prime2},s_{0},s^{\prime}_{0})}{\Pi^{QCD}(M^{2},M^{\prime2},\infty,\infty)}\,\ge\,\frac{1}{2}.
         \end{equation}           
    On the other hand, the lower limits of the Borel parameters, are obtained by requiring the convergence of the OPE series. It means that the perturbative effects receive more contributions than the nonperturbative effects. Moreover, the operators with higher dimensions possess the lower contributions in the OPE series. Hence, we employ the following constraint:
     \begin{equation}
    	R\,(\,M^{2},\,M^{\prime2}\,)=\frac{\Pi^{QCD-dim6}(M^{2},M^{\prime2},s_{0},s^{\prime}_{0})}{\Pi^{QCD}(M^{2},M^{\prime2},s_{0},s_{0}^{\prime})}\,\le\,0.05.
    \end{equation}
      By applying all these conditions, we acquire practical intervals of the Borel parameters as $9\,\mathrm{GeV^{2}}\,\le\,M^{2}\,\le\,12\,\mathrm{GeV^{2}}$ and $6\,\mathrm{GeV^{2}}\,\le\,M^{\prime2}\,\le\,9\,\mathrm{GeV^{2}} $. The continuum thresholds, $s_{0}$ and $s^{\prime}_{0}$, which are introduced by the quark-hadron duality assumption, restrict the upper limits of the integrals in calculations in order to eliminate any contributions to the first excited state and the other higher states and continuum in the initial and final channels. Considering that the sum rules are expected to be stable within the allowed intervals of the Borel parameters, $M^{2}$ and $M^{\prime2}$, the continuum thresholds,  $s_{0}$ and $s^{\prime}_{0}$, are also determined through conditions satisfying this stability. We find     
     \[
     	(\,m_{\Sigma_{b}^{*0}}\,+\,0.3\,)^{2}\,\mathrm{GeV^{2}}\,\le\,s_{0}\,\le\,(\,m_{\Sigma_{b}^{*0}}\,+\,0.5\,)^{2}\,\mathrm{GeV^{2}}, 
     \]
	\begin{equation}
		(\,m_{\Sigma_{c}^{+}}\,+\,0.3\,)^{2}\,\mathrm{GeV^{2}}\,\le\,s_{0}^{\prime}\,\le\,(\,m_{\Sigma_{c}^{+}}\,+\,0.5\,)^{2}\,\mathrm{GeV^{2}}.
	\end{equation}
	As depicted in Figs. \ref{ffm2s0}, \ref{ffm2sp0}, \ref{ffmp2s0}, and \ref{ffmp2sp0}, the form factors have very weak dependence on variations of  $s_{0},\, s^{\prime}_{0},\, M^{2} $, and $M^{\prime2}$ within their practical intervals that can be interpreted as an acceptable consistency between the chosen ranges of the parameters and the requirements of the method.
	
	 Another parameter that is required to be specified, in the QCD sum rules method, is the mixing parameter $\beta$. It is a mathematical parameter that arises from an arbitrary linear combination in interpolating currents of spin-$\frac{1}{2}$ baryons. Therefore, it belongs to an extensive range from $-\infty$ to $+\infty$. In order to restrict the $\beta$ parameter, we define $x\,=\,\cos\theta$, where  $\theta\,=\,\tan^{-1}\beta$. According to the requirements of the method, the working domain of this parameter, $\beta$, is properly chosen in such a way that the behavior of the form factors,  within the selected area, be rather stable. Hence, the practical interval of the $x$ parameter, is determined as
		$-1.0\,\le\,x\,\le\,-0.5$.
	We utilize this condition for computing all the form factors. It is worth mentioning that $x=-0.71$, corresponding to $\beta=-1$ in the Ioffe current, is incorporated in the selected region.

\begin{figure}
	 		\begin{center}
	 			\subfigure{\includegraphics[height=5 cm , width=5.5 cm]{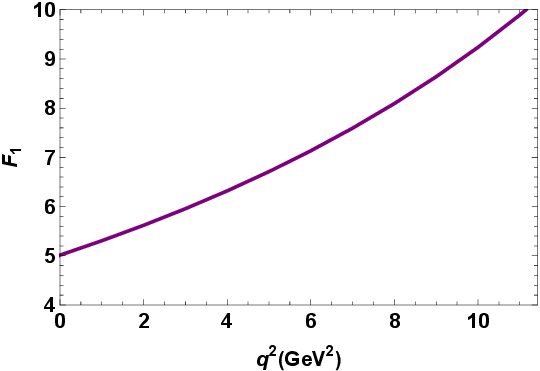}}
	 			\hspace*{0.02 cm}	
	 			\subfigure{\includegraphics[height=5 cm , width=5.5 cm]{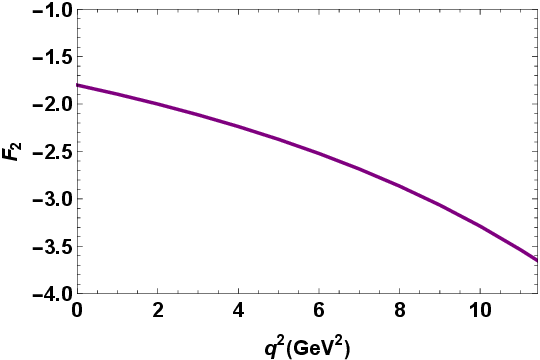}}
	 			\hspace*{0.02 cm}
	 			\subfigure{\includegraphics[height=5 cm , width=5.5 cm]{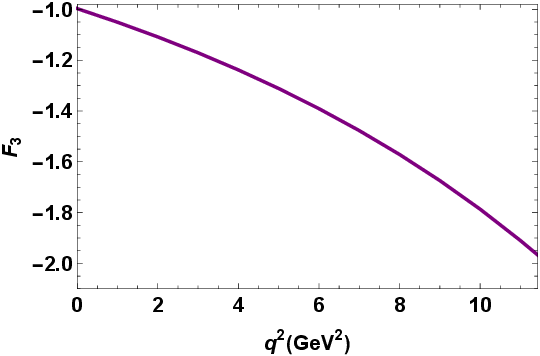}}
	 			
	 			\subfigure{\includegraphics[height=5 cm , width=5.5 cm]{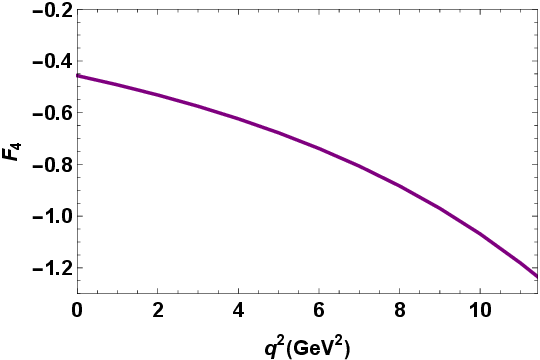}}
	 			\hspace*{0.02 cm}
	 			\subfigure{\includegraphics[height=5 cm , width=5.5 cm]{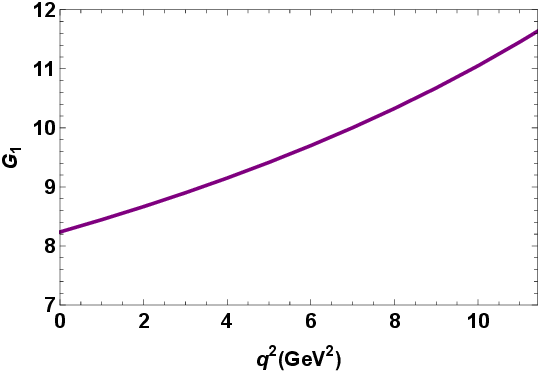}}
	 			\hspace*{0.02 cm}
	 			\subfigure{\includegraphics[height=5 cm , width=5.5 cm]{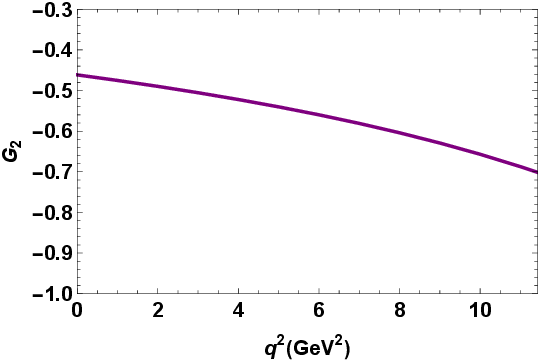}}
	 			
	 			\subfigure{\includegraphics[height=5 cm , width=5.5 cm]{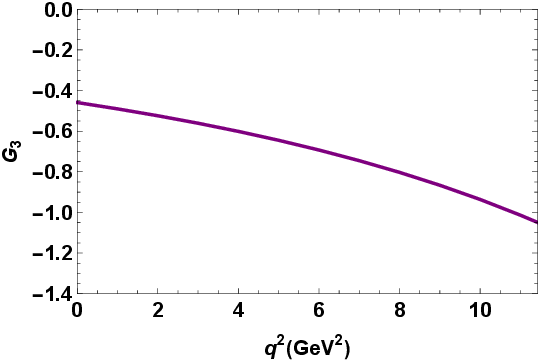}}
	 			\hspace*{0.02 cm}
	 			\subfigure{\includegraphics[height=5 cm , width=5.5 cm]{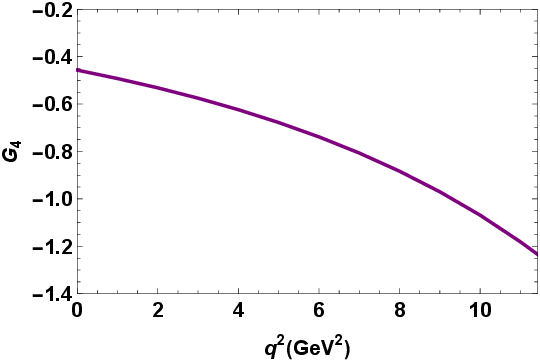}}
	 		\end{center}
	 		\caption{The behavior of the form factors $F_{1},\, F_{2},\, F_{3},\, F_{4},\, G_{1},\, G_{2},\, G_{3}$, and $G_{4}$, represented in Table \ref{set1} with respect to $q^{2}$ at the central values of auxiliary parameters, $ M^{2},\,M^{\prime2},\,s_{0},\,s^{\prime}_{0}$, and Ioffe point.  }
	 		\label{ffq2}
	 	\end{figure}

\begin{figure}
     	\begin{center}
     		\subfigure{\includegraphics[height=5 cm , width=5.5 cm]{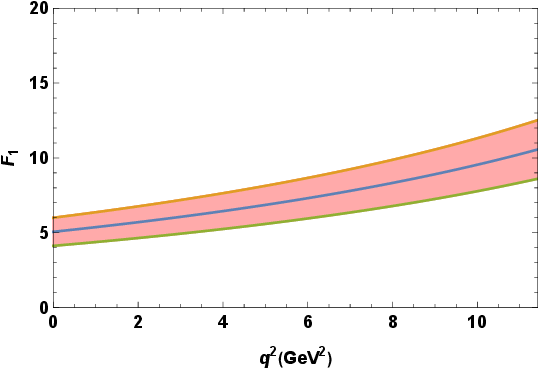}}
     		\hspace*{0.02 cm}	
     		\subfigure{\includegraphics[height=5 cm , width=5.5 cm]{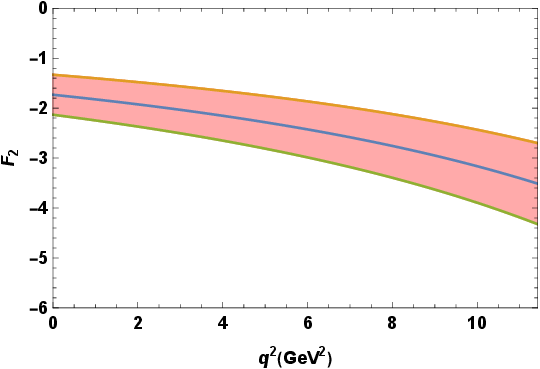}}
     		\hspace*{0.02 cm}
     		\subfigure{\includegraphics[height=5 cm , width=5.5 cm]{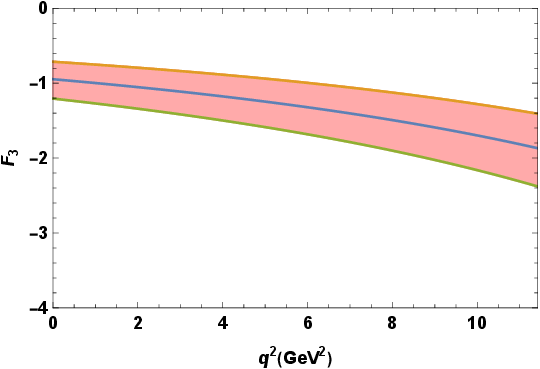}}
     		
     		\subfigure{\includegraphics[height=5 cm , width=5.5 cm]{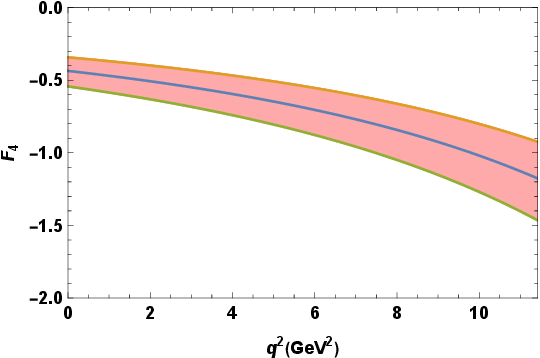}}
     		\hspace*{0.02 cm}
     		\subfigure{\includegraphics[height=5 cm , width=5.5 cm]{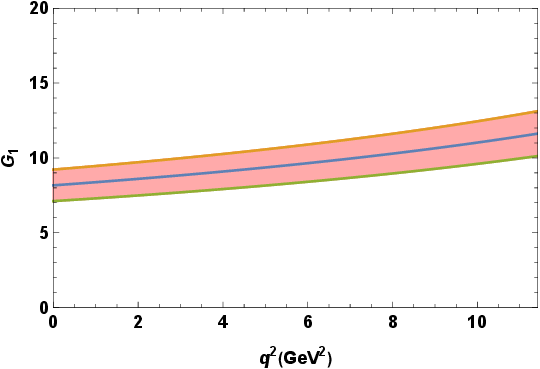}}
     		\hspace*{0.02 cm}
     		\subfigure{\includegraphics[height=5 cm , width=5.5 cm]{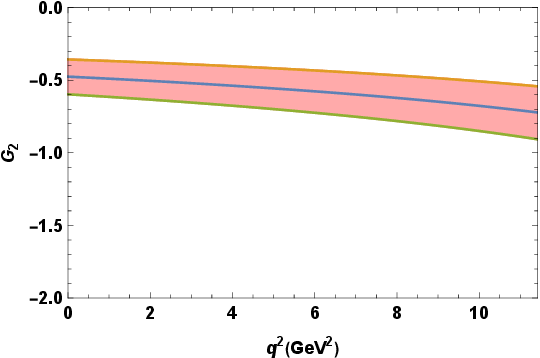}}
     		
     		\subfigure{\includegraphics[height=5 cm , width=5.5 cm]{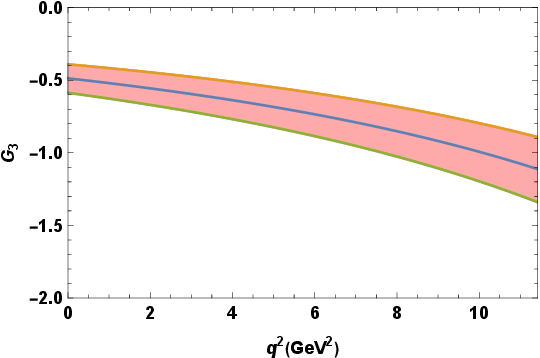}}
     		\hspace*{0.02 cm}
     		\subfigure{\includegraphics[height=5 cm , width=5.5 cm]{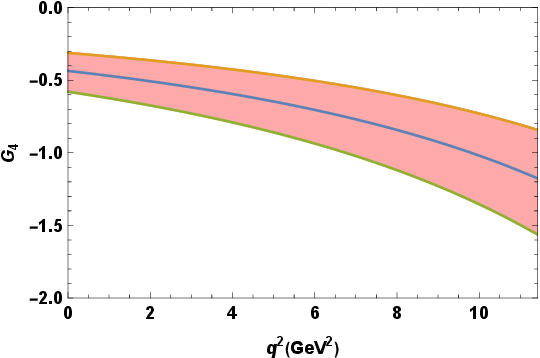}}
     	\end{center}
     	\caption{The error bands of the form factors $F_{1},\, F_{2},\, F_{3},\, F_{4},\, G_{1},\, G_{2},\, G_{3}$ and $G_{4}$, represented in Table \ref{set1} with respect to $q^{2}$ at the center of values of auxiliary parameters, $ M^{2},\,M^{\prime2},\,s_{0},\,s^{\prime}_{0}$ and Ioffe point.}
     	\label{fferror}
     \end{figure}

 		After establishing the proper regions of the auxiliary parameters, we study the behavior of the form factors with respect  to $q^{2}$ that is defined in whole physical region,  $m_{\ell}^{2}\,\le\,q^{2}\,\le (m_{\Sigma_{b}^{*0}}-m_{\Sigma_{c}^{+}})^{2}$. In this region, the maximal value of $q^{2}$ is $q^{2}_{max} = 11.42 \,\mathrm{GeV^{2}}$, and the lowest value of $q^{2}$ is the squared lepton's mass, listed in Table \ref{inputpara}. Following our examination, it is found that the form factors are well fitted to the function as follows:
 	\begin{equation}
 	\mathcal{F}(q^{2})	= \frac{\mathcal{F}(0)}{\bigg(1-a_{1}\,\frac{q^{2}}{m_{\Sigma_{b}^{*0}}^{2}}\,+\,a_{2}\,\frac{q^{4}}{m_{\Sigma_{b}^{*0}}^{4}}\,+\,a_{3}\,\frac{q^{6}}{m_{\Sigma_{b}^{*0}}^{6}}\,+\,a_{4}\,\frac{q^{8}}{m_{\Sigma_{b}^{*0}}^{8}}\bigg)}.
 	\end{equation} 
	 The values of the parameters, $\mathcal{F}(0),\, a_{1},\,a_{2} ,\,a_{3}$, and $a_{4}$,  obtained by employing the central  values of the auxiliary parameters at Ioffe point, $x\,=\,-0.71$, are indicated in Table \ref{set1}.
	  \begin{table}
	 	\centering
	 	\caption{Parameters of the fit functions for the responsible form factors for $\Sigma_{b}^{*0}\rightarrow \Sigma_{c}^{+}\,\ell\,\bar{\nu}_{\ell}$ semileptonic decay.}
	 	\begin{ruledtabular}
	 	\begin{tabular}{|c|c|c|c|c|c|c|c|c|}
	 		&$F_{1}(q^{2})$&$F_{2}(q^{2})$&$F_{3}(q^{2})$&$F_{4}(q^{2})$&$G_{1}(q^{2})$&$G_{2}(q^{2})$&$G_{3}(q^{2})$&$G_{4}(q^{2})$\\
	 		\hline
	 		$\mathcal{F}(q^{2}=0)$&$5.07\pm0.93$&$-1.73\pm0.40$&$-0.95\pm0.14$&$-0.44\pm0.09$&$8.17\pm1.05$&$-0.48\pm0.11$&$-0.49\pm0.09$&$-0.44\pm0.09$\\
	 		$a_{1}$&1.99&1.74&1.76&2.51&0.85&0.97&2.22&2.51\\
	 		$a_{2}$&1.66&0.64&0.92&2.097&-0.12&-0.097&1.91&2.097\\
	 		$a_{3}$&-1.11&0.13&-0.10&-0.575&0.029&-0.086&-0.939&-0.575\\
	 		$a_{4}$&0.51&-0.035&-0.022&-0.023&0.035&-0.0097&0.29&-0.023\\
	 	\end{tabular}
	 \end{ruledtabular}
	 	\label{set1}
	 \end{table}
	   QCD sum rules method for the form factors is based on selecting appropriate structures, which lead to the least possible uncertainties for the form factors by perceiving the practical regions of the Borel parameter,  continuum threshold, and the mixing $x$ parameter.  As it is obvious from the presented results in both the hadronic representation, Eq. \eqref{physical1}, and the QCD representation, Eq. \eqref{QCD1}, the Lorentz structures, in QCD sum rules method, are not unique and the form factors are actually structure dependent.   Hence,  in this study,  we choose the structures $g_{\mu\nu}\slashed{p}\gamma_{5},\,p^{\prime}_{\nu}\gamma_{\mu}\slashed{p}\slashed{p}^{\prime}\gamma_{5},\,p_{\mu}p^{\prime}_{\nu}\slashed{p}\slashed{p}^{\prime}\gamma_{5},\,p^{\prime}_{\mu}p^{\prime}_{\nu}\slashed{p}\slashed{p}^{\prime}\gamma_{5},\,g_{\mu\nu}\slashed{p},\,p^{\prime}_{\mu}p^{\prime}_{\nu}\slashed{p},\,p_{\mu}p^{\prime}_{\nu}\slashed{p}^{\prime}$, and $p^{\prime}_{\mu}p^{\prime}_{\nu}\slashed{p}\slashed{p}^{\prime} $, which show more steadiness and less uncertainties with respect to variations of the helping parameters (for more information  about reducing the uncertainties, see, for instance, Ref.  \cite{Khodjamirian:2011jp}). The selected structures, $g_{\mu\nu}\slashed{p}\gamma_{5},\,p^{\prime}_{\nu}\gamma_{\mu}\slashed{p}\slashed{p}^{\prime}\gamma_{5},\,p_{\mu}p^{\prime}_{\nu}\slashed{p}\slashed{p}^{\prime}\gamma_{5},\,p^{\prime}_{\mu}p^{\prime}_{\nu}\slashed{p}\slashed{p}^{\prime}\gamma_{5},\,g_{\mu\nu}\slashed{p},\,p^{\prime}_{\mu}p^{\prime}_{\nu}\slashed{p},\,p_{\mu}p^{\prime}_{\nu}\slashed{p}^{\prime}$, and $p^{\prime}_{\mu}p^{\prime}_{\nu}\slashed{p}\slashed{p}^{\prime} $, are corresponding to the eight form factors $F_{1},\, F_{2},\, F_{3},\, F_{4},\, G_{1},\, G_{2},\, G_{3}$, and $G_{4}$, respectively, that are represented in Table \ref{set1}.   
	  The uncertainties of the form factors at $q^{2}=0$ originate from the uncertainties appear in determination of the acceptable regions for the auxiliary  parameters and errors of the other input values.
	  The behavior of the form factors  $F_{1},\, F_{2},\, F_{3},\, F_{4},\, G_{1},\, G_{2},\, G_{3}$, and $G_{4}$, corresponding to the structures $g_{\mu\nu}\slashed{p}\gamma_{5},\,p^{\prime}_{\nu}\gamma_{\mu}\slashed{p}\slashed{p}^{\prime}\gamma_{5},\,p_{\mu}p^{\prime}_{\nu}\slashed{p}\slashed{p}^{\prime}\gamma_{5},\,p^{\prime}_{\mu}p^{\prime}_{\nu}\slashed{p}\slashed{p}^{\prime}\gamma_{5},\,g_{\mu\nu}\slashed{p},\,p^{\prime}_{\mu}p^{\prime}_{\nu}\slashed{p},\,p_{\mu}p^{\prime}_{\nu}\slashed{p}^{\prime}$, and $p^{\prime}_{\mu}p^{\prime}_{\nu}\slashed{p}\slashed{p}^{\prime} $, as a function of $q^{2}$ at  the central values of $ M^{2},\,M^{\prime2},\,s_{0},\,s^{\prime}_{0}$, and the Ioffe point ($x=-0.71$), is illustrated in Figs. \ref{ffq2} and \ref{fferror}, without and with considering uncertainties, respectively. As Fig. \ref{ffq2} depicts, all the form factors, related to the weak transitions, increase by raising $q^{2}$ that is consistent with the expectations. In the next section, we employ the fit functions of the form factors in the physical region $m_{\ell}^{2}\,\le\,q^{2}\,\le (m_{\Sigma_{b}^{*0}}-m_{\Sigma_{c}^{+}})^{2}$ to compute the corresponding decay widths in all lepton channels.
	           
	 \section{COMPUTATION OF DECAY WIDTH}\label{width}
	 The acquired results for the fit functions of the form factors, in the previous section,  enable us to evaluate the decay widths of the $\Sigma_{b}^{*0} \rightarrow \Sigma_{c}^{+}\,\ell\,\bar{\nu}_{\ell}$ transitions in all lepton channels. To this end, we compute vector and axial vector helicity amplitudes, $ H^{V,A}_{\lambda_{2},\lambda_{W}}$, represented in respect of linear expressions of the vector, $F_{i}^{V} \equiv F_{i} (i=1,2,3,4)$, and axial vector, $F_{i}^{A}\equiv G_{i}$, invariant transition form factors, where $\lambda_{W}=t,\,0,\,\pm 1$ and $\lambda_{2}=\pm \frac{1}{2}\,$ are the helicity components of the virtual W boson and the final baryon, respectively \cite{Gutsche:2018nks, Faessler:2009xn}. The helicity of the initial baryon, $ \lambda_{1}$, is established by the relation $ \lambda_{1}=\lambda_{2}-\lambda_{W} $. Vector and axial vector parts of helicity amplitudes are determined by using $ H^{V,A}_{\lambda_{2},\lambda_{W}}\,=\,M_{\mu}^{V,A}\,\bar{\epsilon}^{*\mu}(\lambda_{W})$, where $M_{\mu}^{V,A}$, as expressed in Eq. \eqref{formfactor}, are the vector and axial vector currents, $J_{\mu}^{V,A}$, which are placed between the initial and final baryonic states and $\bar{\epsilon}^{*\mu}$ is the polarization vector of W boson. Hence, we have  
	 \begin{eqnarray}
	 	&H^{V}_{\frac{1}{2},t}\,=\,-\sqrt{\frac{2}{3}}\,\alpha^{V}_{\frac{1}{2},t}\,(\omega-1)\,\Big(\,F_{1}^{V}M_{2}+F_{2}^{V}M_{+}-F_{3}^{V}\frac{M_{1}}{M_{2}}(M_{1}-M_{2}\,\omega)-F_{4}^{V}\frac{q^{2}}{M_{2}}\Big),\nonumber\\
	 	&H^{V}_{\frac{1}{2},0}\,=\,-\sqrt{\frac{2}{3}}\,\alpha^{V}_{\frac{1}{2},0}\,\Big(\,F_{1}^{V}(M_{1}-M_{2}\,\omega)\,+\,F_{2}^{V}(\omega+1)M_{-}-F_{3}^{V}(\omega^{2}-1)M_{1}\Big),\nonumber\\
	 	&H^{V}_{\frac{1}{2},1}\,=\,\frac{1}{\sqrt{6}}\,\alpha^{V}_{\frac{1}{2},1}\,\Big(\,F_{1}^{V}\,+\,2\,F_{2}^{V}\,(\omega+1)\Big),\nonumber\\
	 	&H^{V}_{\frac{1}{2},-1}\,=\,\frac{1}{\sqrt{2}}\,\alpha^{V}_{\frac{1}{2},1}\,F_{1}^{V},
	 	\end{eqnarray}
	 	and
	 	\begin{eqnarray}
	 	&H^{A}_{\frac{1}{2},t}\,=\,\sqrt{\frac{2}{3}}\,\alpha^{A}_{\frac{1}{2},t}\,(\omega+1)\,\Big(\,F_{1}^{A}M_{2}-F_{2}^{A}M_{-}-F_{3}^{A}\,\frac{M_{1}}{M_{2}}(M_{1}-M_{2}\,\omega)-F_{4}^{A}\frac{q^{2}}{M_{2}}\Big),\nonumber\\
	 	&H^{A}_{\frac{1}{2},0}\,=\,-\sqrt{\frac{2}{3}}\,\alpha^{A}_{\frac{1}{2},0}\,\Big(\,-F_{1}^{A}(M_{1}-M_{2}\,\omega)\,+\,F_{2}^{A}(\omega-1)M_{+}+F_{3}^{A}(\omega^{2}-1)M_{1}\Big),\nonumber\\
	 	&H^{A}_{\frac{1}{2},1}\,=\,\frac{1}{\sqrt{6}}\,\alpha^{A}_{\frac{1}{2},1}\,\Big(\,-F_{1}^{A}\,+\,2\,F_{2}^{A}\,(\omega-1)\Big),\nonumber\\
	 	&H^{A}_{\frac{1}{2},-1}\,=\,-\frac{1}{\sqrt{2}}\,\alpha^{A}_{\frac{1}{2},1}\,F_{1}^{A},
	 \end{eqnarray}
	  where the following definitions are utilized:
	 \begin{eqnarray}
	 	&&\alpha^{V}_{\frac{1}{2},t}\,=\,\alpha^{A}_{\frac{1}{2},0}\,=\,\sqrt{\frac{2M_{1}M_{2}(\omega+1)}{q^{2}}},\nonumber\\
	 	&&\alpha^{V}_{\frac{1}{2},0}\,=\,\alpha^{A}_{\frac{1}{2},t}\,=\,\sqrt{\frac{2M_{1}M_{2}(\omega-1)}{q^{2}}},\nonumber\\
	 	&&\alpha^{V}_{\frac{1}{2},1}\,=\,2\,\sqrt{M_{1}M_{2}(\omega-1)},\nonumber\\
	 	&&\alpha^{A}_{\frac{1}{2},1}\,=\,2\,\sqrt{M_{1}M_{2}(\omega+1)},
	 \end{eqnarray}
	 with
	 \[
	 M_{\pm}\,=\,M_{1}\pm M_{2}, \,\,\,\,\,\,\,\,\omega\,=\,\frac{M_{1}^{2}+M_{2}^{2}-q^{2}}{2M_{1}M_{2}},
	 \]
     in which $M_{1}$ and $M_{2}$ are the masses of the initial, $\Sigma_{b}^{*0}$, and final, $\Sigma_{c}^{+}$, baryons, respectively. The negative helicity amplitudes are expressed as
     \begin{equation}
     	H^{V}_{-\lambda_{2},-\lambda_{W}}\,=\,-	H^{V}_{\lambda_{2},\lambda_{W}}\,\,\,\, ,\,\,\,\,H^{A}_{-\lambda_{2},-\lambda_{W}}\,=\,	H^{A}_{\lambda_{2},\lambda_{W}}.
     \end{equation} 
    Ultimately, the helicity amplitude for the V-A current in weak transitions is given as $H_{\lambda_{2},\lambda_{W}}\,=\,H^{V}_{\lambda_{2},\lambda_{W}}\,-\,H^{A}_{\lambda_{2},\lambda_{W}}$. 
      Having obtained the helicity amplitudes, the decay width for the semileptonic $\Sigma_{b}^{*0} \rightarrow \Sigma_{c}^{+}\,\ell\,\bar{\nu}_{\ell}$ transition is computed by the following formula \cite{Gutsche:2018nks, Faessler:2009xn}:
      \begin{equation}
      	\Gamma_{\Sigma_{b}^{*0}\rightarrow \Sigma_{c}^{+}\, \ell \,\bar{\nu}_{\ell}}\,=\,\frac{1}{2}\,\frac{G_{F}^{2}\,|V_{cb}|^{2}}{192\,\pi^{3}}\,\frac{M_{\Sigma_{c}^{+}}}{M_{\Sigma_{b}^{*0}}^{2}}\,\int_{m_{\ell}^{2}}^{(M_{\Sigma_{b}^{*0}}-M_{\Sigma_{c}^{+}})^{2}}\,\frac{\mathrm{d}q^{2}}{q^{2}}\,(q^{2}-m_{\ell}^{2})^{2}\,\sqrt{\omega^{2}-1}\,\,\mathcal{H}_{3/2 \rightarrow 1/2},
      \end{equation}
     where $m_{\ell}$ represents the lepton mass and $\mathcal{H}_{3/2 \rightarrow 1/2}$ includes combinations of the helicity amplitudes,
     \begin{eqnarray}
   &&  \mathcal{H}_{3/2 \rightarrow 1/2}\,=\,\lvert H_{\frac{1}{2},1}\rvert^{2}\,+\,\lvert H_{-\frac{1}{2},-1}\rvert^{2}\,+\,\lvert H_{\frac{1}{2},-1}\rvert^{2}\,+\,\lvert H_{-\frac{1}{2},1}\rvert^{2}\,+\,\lvert H_{\frac{1}{2},0}\rvert^{2}\,+\,\lvert H_{-\frac{1}{2},0}\rvert^{2}\,\nonumber\\
     &&+\,\frac{m_{\ell}^{2}}{2q^{2}}\,\bigg(3\lvert H_{\frac{1}{2},t}\rvert^{2}\,+3\lvert H_{-\frac{1}{2},t}\rvert^{2}\,+\lvert H_{\frac{1}{2},1}\rvert^{2}\,+\lvert H_{-\frac{1}{2},-1}\rvert^{2}\,+\lvert H_{\frac{1}{2},-1}\rvert^{2}\,+\lvert H_{-\frac{1}{2},1}\rvert^{2}\,+\lvert H_{\frac{1}{2},0}\rvert^{2}\,+\lvert H_{-\frac{1}{2},0}\rvert^{2}\bigg).
     \end{eqnarray}
     The obtained results are presented in Table \ref{width20}. Our findings can be examined by upcoming experimental data and they can provide us with useful information about the inner structures of the $\Sigma_{b}^{*0}$ and $\Sigma_{c}^{+}$ baryons.
      \begin{table}[h]
     	\centering
     	\caption{Decay widths of the semileptonic  $\Sigma_{b}^{*0}\rightarrow \Sigma_{c}^{+}\,\ell\,\bar{\nu}_{\ell}$ transition at different lepton channels in units of GeV.}
     	\begin{ruledtabular}
     	\begin{tabular}{|c|c|c|}
     		$ $& & \\
     		$\Gamma[\,\Sigma_{b}^{*0}\rightarrow \Sigma_{c}^{+}\,e\,\bar{\nu}_{e}\,]\times 10^{12}\,\,\,$&$\Gamma[\,\Sigma_{b}^{*0}\rightarrow \Sigma_{c}^{+}\,\mu\,\bar{\nu}_{\mu}\,]\times 10^{12}\,\,\,\,\,\,\,\,\,\,\,$&$\Gamma[\,\Sigma_{b}^{*0}\rightarrow \Sigma_{c}^{+}\,\tau\,\bar{\nu}_{\tau}\,]\times 10^{13}\,\,\,\,\,\,\,\,\,\,\,$\\
     		$ $& & \\
     		\hline
     		$1.34^{+0.38}_{-0.33}\,\,\,\,$&$1.33^{+0.38}_{-0.33}\,\,\,\,\,\,\,\,\,\,\,$&$3.85^{+1.08}_{-0.95}\,\,\,\,\,\,\,\,\,\,\,$\\
     	\end{tabular}
     	\end{ruledtabular}
     	\label{width20}
     \end{table}
     Decay width, as an observable quantity of considerable importance, can be employed to exhibit the dependency associated with the lepton channel. Indeed, the decay width decreases by raising lepton mass that is consistent with the kinematical constraints. It is convenient to evaluate the ratio of decay widths since a number of experimental systematic uncertainties will be eliminated.  Hence,  we calculate the ratio of decay width in $ \tau $  to $e/\mu$ channel as follows:
    \begin{eqnarray}\label{Ratio}
       R\,=\,\frac{\Gamma[\,\Sigma_{b}^{*0}\rightarrow \Sigma_{c}^{+}\,\tau\,\bar{\nu}_{\tau}\,]}{\Gamma[\,\Sigma_{b}^{*0}\rightarrow \Sigma_{c}^{+}\,e\,(\mu)\,\bar{\nu}_{e\,(\mu)}\,]}\,=\,0.29\pm 0.01.
     \end{eqnarray}
      These results may assist future experiments in searching for these  decay channels and test the SM prediction. Any sizable deviations of the future data from our SM predictions can be consider as a sign for new physics BSM.
	 \section{conclusion}\label{clue}
	Probing  structures and properties  of the heavy baryons is one of the main aims of various experiments. Consequently, the study of heavy baryons from different aspects, especially decay properties, has triggered considerable theoretical interest. The weak decays provide noteworthy information on the structures of the heavy baryons, and they have been in  concentration of various theoretical studies. In the present work, we explored the b-heavy baryon $\Sigma_{b}^{*}$. Its strong and radiative decay modes have been previously analyzed \cite{Kim:2022pyq,ParticleDataGroup:2024cfk}; however, there are few detailed analyses of its weak decay channels. Hence, we investigated the semileptonic weak decay of this singly heavy baryon, $\Sigma_{b}^{*0}$, with spin $\frac{3}{2}$, into the singly heavy baryon, $\Sigma_{c}^{+}$, with spin $\frac{1}{2}$ in the $\Sigma_{b}^{*0}\rightarrow \Sigma_{c}^{+}\, \ell \,\bar{\nu}_{\ell}$ transition for three lepton channels within the QCD sum rule method. Our computations include both the perturbative and nonperturbative contributions of the OPE series up to mass dimension six. We utilized the interpolating currents of the singly heavy baryons $\Sigma_{b}^{*0}$ and $\Sigma_{c}^{+}$ as the initial and final states, respectively, to produce the sum rules for the semileptonic weak form factors. After fixing the auxiliary parameters of the method, we acquired the fit functions of the eight responsible form factors in terms of $q^{2}$ in the whole kinematic range accessible in semileptonic decays, $m_{\ell}^{2}\,\le\,q^{2}\,\le (m_{\Sigma_{b}^{*0}}-m_{\Sigma_{c}^{+}})^{2}$. The obtained results were employed to evaluate the decay widths in all lepton channels.
	 Regarding the significant progress made at various experiments, it is widely anticipated establishing many more rare decay channels in future \cite{Kholodenko:2024bjr}. The present  pioneering research on the considered weak decays can support/guide various related experiments and lead to the identification  of new decay modes.
	 %%%%%%%%%%%%%%%%%%%%%%%%%%%%%%%%%%%%%%%%%%%%%%%%%%%%%%%%%%%%%%%%%%%%%%%%%%%%%%%%%%%%%%%
	 
	 As final remark, we would like to provide some information may  help experimental groups aiming  to study the properties of $\Sigma_{b}^{*}$ baryon or even  the decay mode $\Sigma_{b}^{*}\,\rightarrow\, \Sigma_{c}\,\ell\,\bar{\nu}_{\ell}$,  investigated  in the present work,  by  considering the great new advances in experimental facilities. As  previously stated,  the $\Sigma_{b}^{*}$ baryons were produced at the CDF Collaboration in the   $\Lambda_{b}^{0}\,\pi^{\pm}$ invariant mass at an integrated luminosity of $ 1.1 \,fb^{-1} $  and $ \sqrt{s}=1.96\, TeV $ in $ p\bar{p} $ collision \cite{CDF:2007oeq}. Among the four produced spin $\frac{1}{2}  $ and  $\frac{3}{2}  $ 
	 $\Sigma_{b}$ states, the average central yield for spin  $\frac{3}{2}  $  states was roughly $ 73 $ compared to the spin  $\frac{1}{2}  $ resonances with $ 46 $ yield. Considering the  LHCb experiment, as an example,  with integrated luminosity of $ 9.56 \,fb^{-1} $  and $ \sqrt{s}=13.6\, TeV $ \cite{LHCb}, and the fact that the number of produced events is directly proportional to the integrated luminosity and cross section, it is expected more $\Sigma_{b}^{*}$ states to be produced in the present and future runs of the LHC.  One may have some approximate estimations on the number of events that are produced/to be produced at the present and future runs of the LHC compared to the CDF Collaboration considering the same cross section. As it is seen, the average yield with the present condition of the  LHCb 
	 experiment is  roughly 10 times greater than that of the CDF Collaboration, i.e. $ 730 $ events. The LHCb experiment aims to collect a dataset of $ 300 \,fb^{-1} $  in its high-luminosity phase \cite{Manuzzi:2025arg}. Considering such a huge luminosity, we expect roughly $ 21900 $ $\Sigma_{b}^{*}$ states. By the help of newly developed machine learning techniques such as generative adversarial networks (GANs) and conditional GANs to augment the dataset, we hope we can easily study  $\Sigma_{b}^{*}$ baryon and its 
	 decay channels in near future.
	 
	 The dominant decay channel of $\Sigma_{b}^{*}$  is  $\Lambda_{b}^{0}\,\pi$ at  which the  discovery was made \cite{ParticleDataGroup:2024cfk}. However, considering the weak tree-level decay channels of this state, which have more contribution compared to the loop-level processes, and the value of $ V_{cb} $ as the element of CKM matrix, which is 11 times greater than the $ V_{ub} $ element, we expect that the $ b\rightarrow c $  processes to have more contributions to the total width compared to the $ b\rightarrow u $ ones as other possible dissociation modes of $\Sigma_{b}^{*}$  state at quark level.  This statement becomes more strong by considering the fact  that the decay rate is proportional to the square  of the CKM matrix elements. We have another fact that, during the time passed, the baryons generally  have discovered from the lightest to the heaviest and from the lower spin to the higher spin. This is the case for detection of their decay channels as well. Comparing the order of decay rate for $\Sigma_{b}^{*}\,\rightarrow\, \Sigma_{c}\,\ell\,\bar{\nu}_{\ell}$ considered in the present work with the $\Lambda_{b} \,\rightarrow\, \Lambda_{c}\,\ell\,\bar{\nu}_{\ell}$ studied with the same method in Ref. \cite{Azizi:2018axf}, we see that the decay rate in our case is roughly $10^{2}$ times greater than the latter. The $\Lambda_{b} \,\rightarrow\, \Lambda_{c}\,\ell\,\bar{\nu}_{\ell}$ processes have been already seen in the experiment in all lepton channels \cite{ParticleDataGroup:2024cfk}. Although, the number of $ \Lambda_{b} $ baryons that are produced at different experiments are very high compared to other b baryons, we expect that we will witness of observing semileptonic decay channels of other members of  the $\frac{1}{2}  $ and $\frac{3}{2}  $ single heavy baryons by the recent experimental progresses made. Considering the information above on the rate of $\Sigma_{b}^{*}$ production in the present and future experiments, we expect to detect the process $\Sigma_{b}^{*}\,\rightarrow\, \Sigma_{c}\,\ell\,\bar{\nu}_{\ell}$ and measure its parameters. It is possible to have a rough estimation for the number of  $\Sigma_{b}^{*}\,\rightarrow\, \Sigma_{c}\,\ell\,\bar{\nu}_{\ell}$ process that would be produced, for instance, at high-luminosity phase of LHCb experiment with aforementioned conditions, in near future. As previously said, the $\Lambda_{b}^{0}\,\pi$ is not only the dominant decay channel of $\Sigma_{b}^{*}$ baryon, but also for the  $\Sigma_{b}$ state as well. Unfortunately, we have no enough information about the exact contribution of $\Sigma_{b}^{*}\,\rightarrow\,\Lambda_{b}^{0}\,\pi$ to the total width of $\Sigma_{b}^{*}$ state, but comparing the result obtained for the width of  $\Sigma_{b}\,\rightarrow\,\Lambda_{b}^{0}\,\pi$ in Ref. \cite{Azizi:2008ui} and total width of $\Sigma_{b}$ baryon presented in the particle data group (PDG) \cite{ParticleDataGroup:2024cfk}, we see a $76\%$ contribution when the central values are considered. If we consider the same amount of contribution for the $\Sigma_{b}^{*}\,\rightarrow\,\Lambda_{b}^{0}\,\pi$ decay channel from the total width of the $\Sigma_{b}^{*}$ state, there remains roughly  $24\%$ contribution to the weak semileptonic and nonleptonic decays of $\Sigma_{b}^{*}$ baryon. Looking at the b-baryon decays in PDG, especially  the $\Lambda_{b}$ baryon modes, we see a $8.55:1$ ratio for the semileptonic to nonleptonic decay modes. As we previously mentioned, the main contribution belongs to the $b \rightarrow c $ based semileptonic transitions. By an approximate calculation, $4706$ events would be expected for the $\Sigma_{b}^{*}\,\rightarrow\, \Sigma_{c}\,\ell\,\bar{\nu}_{\ell}$ process in high luminosity phase of LHCb. 
	  The ratio given in Eq. (\ref{Ratio}) or the ratio of the weak  semileptonic to weak nonleptonic decays  will be relatively easy to access in the experiments. We hope, our results help experimental groups in the course of their search for such decay channels.
	 %%%%%%%%%%%%%%%%%%%%%%%%%%%%%%%%%%%%%%%%%%%%%%%%%%%%%%%%%%%%%%%%%%%%%%%%%%%%%%%%%%

	 	\section*{ACKNOWLEDGMENTS}
	 K. Azizi is grateful to the CERN-TH division for their support and warm hospitality.
	 
	 \section*{DATA AVAILABILITY}
	 No data were created or analyzed in this study.
	 
	 \appendix
	 \section{THE INTERPOLATING CURRENT OF SPIN-$\frac{3}{2}$ BARYON}  \label{appA}
	  Interpolating currents, as the primary inputs in the QCD sum rule method, are associated with the properties of the baryons and expressed in terms of quark contents. Derivation of the interpolating current for the $\Sigma_{Q}^{*}$, the single heavy baryon with $Q=b$ or $c$ heavy quark and $J^{p}=\frac{3}{2}^{+}$ as the spin-parity of this state, is discussed in this appendix. First, we consider the construction of a diquark structure with spin 1 and then the third quark with spin $\frac{1}{2}$ will be attached to it. The structure of a diquark is similar to the structure of a meson that is of the form, $J_{meson}\,=\,\bar{q}\,\Gamma\,q$,
	  where  $\Gamma\,=\,\lbrace \,I\, , \,\gamma_{5}\, ,\,\gamma_{\mu}\, ,\gamma_{\mu}\gamma_{5}\, ,\sigma_{\mu\nu} \,\rbrace $ is Dirac set and $q$ represents the light quark spinor. By replacing the antiquark with its charge conjugation analog, $\bar{q}\,=\,q^{T}\,C$, the structure of diquark will be obtained as $J_{diquark}\,=\,q^{T}\,C\,\Gamma\,q$,
	  where $C\,=i\gamma_{2}\,\gamma_{0}$ is the charge conjugation operator. Attaching the third quark and considering the color indices, $(a, b$, and $c)$, the interpolating current has the general form as
	  \begin{equation}\label{current32}
	  	\mathcal{J}^{\Sigma_{Q}^{*}}\,\sim\,\epsilon^{abc}\,\big\lbrace(\,q^{aT}\,C\,\Gamma\,q^{\prime b}\,)\,\Gamma^{\prime}\,Q^{c}+(\,q^{aT}\,C\,\Gamma\,Q^{b}\,)\,\Gamma^{\prime}\,q^{\prime c}+(\,q^{\prime aT}\,C\,\Gamma\,Q^{b}\,)\,\Gamma^{\prime}\,q^{c}\big\rbrace,
	  \end{equation} 	 
	 Considering all quantum numbers and requirements, $\Gamma$ and $\Gamma^{\prime}$ will be determined. $\epsilon^{abc}$ is the Levi-Civita tensor that makes the interpolating current, color singlet. In the first part of \eqref{current32}, $(\,q^{aT}\,C\,\Gamma\,q^{\prime b}\,)\,\Gamma^{\prime}\,Q^{c}$, since the diquark has spin 1, it should be symmetric under the exchange of the two light quarks  $q\,\leftrightarrow\,q^{\prime}$. Transposing leads to
	 \begin{equation}
	 	 [\,\epsilon_{abc}\,(\,q^{aT}\,C\,\Gamma\,q^{\prime b}\,)]^{T}\,=\,-\epsilon_{abc}\,q^{\prime b T}\,\Gamma^{T}\,C^{-1}\,q^{a}\,=\,\epsilon_{abc}\,q^{\prime b T}\,C\,(C\,\Gamma^{T}\,C^{-1})\,q^{a} ,
	 \end{equation}
	 where we consider the fact that the quark fields are Grassmann numbers and utilize the identities  $C^{T}\,=\,C^{-1}$ and $C^{2}\,=\,-1$. Therefore, we obtain
	 \begin{equation}
	  C\,\Gamma^{T}\,C^{-1}\,=\,
	 \begin{cases}
	 	\Gamma & \text{for} \,\,\,\,\Gamma =1,\gamma_{5},\gamma_{\mu}\gamma_{5}, \\
	 	-\Gamma & \text{for} \,\,\,\,\Gamma=\gamma_{\mu},\sigma_{\mu\nu}.
	 \end{cases}
	 \end{equation} 
	 After switching color dummy indices, we acquire 
	 \begin{equation}\label{twoqq}
	 	 [\,\epsilon_{abc}\,(\,q^{aT}\,C\,\Gamma\,q^{\prime b}\,)]^{T}\,=\,\pm\,\epsilon_{abc}\,q^{\prime a T}\,C\,\Gamma\,q^{b},
	  \end{equation}	
	 where $+$ sign belongs to $\Gamma=\gamma_{\mu},\sigma_{\mu\nu}$ and $-$ sign is related to $\Gamma =1,\gamma_{5},\gamma_{\mu}\gamma_{5}$. By applying the condition of being symmetric under the exchange of light quarks to the right-hand side of the Eq. \eqref{twoqq}, we have
	 \begin{equation}
	 	 [\,\epsilon_{abc}\,(\,q^{aT}\,C\,\Gamma\,q^{\prime b}\,)]^{T}\,=\,\pm\,\epsilon_{abc}\,q^{a T}\,C\,\Gamma\,q^{\prime b},
	 	\end{equation}
	 where the $+$ and $-$ signs are for $\Gamma=\gamma_{\mu},\sigma_{\mu\nu}$ and $\Gamma =1,\gamma_{5},\gamma_{\mu}\gamma_{5}$, respectively. Considering that the transposition of a $(1\times 1)$ matrix like $\epsilon_{abc}\,q^{a T}\,C\,\Gamma\,q^{\prime b}$ should be equal to itself, we determine the only possible choice for $\Gamma$ as $ \Gamma\,=\,\gamma_{\mu} $ or $\sigma_{\mu\nu}$. These conditions should be applied to the  other structures, $(\,q^{aT}\,C\,\Gamma\,Q^{b}\,)\,\Gamma^{\prime}\,q^{\prime c}$ and $(\,q^{\prime aT}\,C\,\Gamma\,Q^{b}\,)\,\Gamma^{\prime}\,q^{c}$, of \eqref{current32}. Hence we have two possible forms of the interpolating current as 
	 \begin{eqnarray}
	 	 &&\epsilon_{abc}\,\Big\lbrace\, (\,q^{aT}\,C\,\gamma_{\mu}\,q^{\prime b\,})\,\Gamma_{1}^{\prime}\,Q^{c}\,+\,
	 	(\,q^{aT}\,C\,\gamma_{\mu}\,Q^{b\,})\,\Gamma_{1}^{\prime}\,q^{\prime c}\,+\,
	 	(\,Q^{aT}\,C\,\gamma_{\mu}\,q^{\prime b\,})\,\Gamma_{1}^{\prime}\,q^{c}\,\Big \rbrace ,\notag\\
	 	&&\epsilon_{abc}\,\Big\lbrace\, (\,q^{aT}\,C\,\sigma_{\mu\nu}\,q^{\prime b\,})\,\Gamma_{2}^{\prime}\,Q^{c}\,+\,
	 	(\,q^{aT}\,C\,\sigma_{\mu\nu}\,Q^{b\,})\,\Gamma_{2}^{\prime}\,q^{\prime c}\,+\,
	 	(\,Q^{aT}\,C\,\sigma_{\mu\nu}\,q^{\prime b\,})\,\Gamma_{2}^{\prime}\,q^{c}\,\Big \rbrace.
	 \end{eqnarray}
	 By employing the Lorentz and parity considerations, $\Gamma_{1}^{\prime}$ and $\Gamma_{2}^{\prime}$ are determined. The interpolating current of the state with $J=\frac{3}{2}$ has the Lorentz vector structure, therefore, $\Gamma_{1}^{\prime}\,=\,1\,\,\text{or}\,\,\gamma_{5}$ and $\Gamma_{2}^{\prime}\,=\,\gamma_{\nu}\,\,\text{or}\,\,\gamma_{\nu}\gamma_{5}$.
	 After applying the parity transformation, we determine $\Gamma_{1}^{\prime}$ and $\Gamma_{2}^{\prime}$ as $\Gamma_{1}^{\prime}\,=\,1$ and $\Gamma_{2}^{\prime}\,=\,\gamma_{\nu}$. Ultimately, considering all requirements and Fierz identity, we acquire the following form of the interpolating current for a spin-$\frac{3}{2}$ baryon;
	 \begin{equation}
	 	\epsilon_{abc}\,\Big\lbrace\, (\,q^{aT}\,C\,\gamma_{\mu}\,q^{\prime b\,})\,Q^{c}\,+\,
	 	(\,q^{aT}\,C\,\gamma_{\mu}\,Q^{b\,})\,q^{\prime c}\,+\,
	 	(\,Q^{aT}\,C\,\gamma_{\mu}\,q^{\prime b\,})\,q^{c}\,\Big \rbrace.
	 \end{equation} 
	 Particularly, after normalization, for $\Sigma_{b}^{*0}$ we have
	 \begin{equation}
	 		{\cal J}_{\mu}^{\Sigma_{b}^{*0}}(x)\,=\,\sqrt\frac{2}{3}\,\epsilon^{abc}\Bigg\lbrace \Big(u^{aT}(x)C\gamma_{\mu}\,d^{b}(x)\,\Big)b^{c}(x)\,+\,\Big(d^{aT}(x)C\gamma_{\mu}\,b^{b}(x)\,\Big)u^{c}(x)\,+\,\Big(b^{aT}(x)C\gamma_{\mu}\,u^{b}(x)\,\Big)d^{c}(x)\,\Bigg\rbrace.
	 	 \end{equation}
	 	 \section{THE EXPRESSIONS‌ OF THE‌ PERTURBATIVE AND NONPERTURBATIVE CONTRIBUTIONS }\label{appB}
	In this appendix, we express the explicit forms of the spectral densities, $\rho_{i}(s,s^{\prime},q^{2})$, and functions $\Gamma_{i}(p^{2},p^{\prime2},q^{2})$ as an example for the structure $g_{\mu\nu}\,\slashed p$,
		\begin{eqnarray}
		&&\rho_{g_{\mu\nu}\,\slashed p}^{Pert.}(s,s^{\prime},q^{2})\,=\,\int_{0}^{1}\mathrm{d}u\,\int_{0}^{1-u}\mathrm{d}v\,\frac{1}{512\,\sqrt{3}\,\pi^{4}\,F^{2}}\,\notag\\&&\Bigg\lbrace\,-\Bigg(F\,\bigg(m_{b}u\Big(8u\big((1+2\beta)\,(q^{2}-s-s^{\prime})+4\beta s\,u\big)+\Big(-\big((13+32\beta)\,s^{\prime}\big)+32\beta(-q^{2}+s+s^{\prime})u\Big)v+32\beta s^{\prime}v^{2}\Big)\notag\\&&+m_{c}\,\Big(32\beta s(-1+u)\,u^{2}+8(-q^{2}+s+s^{\prime})\,u\,\big(1+\beta(-2+4u)\big)v+s^{\prime}(13+32\beta u)v^{2}\Big)\bigg)\Bigg)\notag\\&&+L\,\Big(5m_{c}v+16\beta m_{c}\big(-v+4uF\big)+m_{b}u\big(-5+16\beta(-5+6u+6v)\big)\Big)\Bigg\rbrace\,\Theta[L(s,s^{\prime},q^{2})],
	\end{eqnarray} 
	\begin{eqnarray}
		\rho_{g_{\mu\nu}\,\slashed p}^{Dim-3}\,(s,s^{\prime},q^{2})=\,\int_{0}^{1}\,\mathrm{d}u\,\int_{0}^{1-u}\,\mathrm{d}v\,\frac{-1}{128\,\sqrt{3}\,\pi^{2}}\,\big(\langle\,\bar{u}u\,\rangle\,+\,\langle\,\bar{d}d\,\rangle\big)\,\big(\,-37\,+\,48\,u\,+\,16\,\beta\,(-3+7\,u)\,\big)\,\Theta[L(s,s^{\prime},q^{2})],
	\end{eqnarray}
	\begin{equation}
		\rho_{g_{\mu\nu}\,\slashed{p}}^{Dim-4}(s,s^{\prime},q^{2})\,=\,0,
	\end{equation}
	
	\begin{equation}
		\rho_{g_{\mu\nu}\,\slashed{p}}^{Dim-5}(s,s^{\prime},q^{2})\,=\,0,
	\end{equation}
	
		\begin{equation}
		\Gamma_{g_{\mu\nu}\,\slashed{p}}^{Dim-6}(p^{2},p^{\prime2},q^{2})\,=\,\frac{1}{3\,\sqrt{3}}\,\frac{1}{B\,B^{\prime}}\,m_{c}\,\langle \bar{u}\,u\,\rangle\,\langle \bar{d}\,d\,\rangle\,(\,-1\,+\,\beta\,),
	\end{equation}
	where,
	\begin{equation}
	L(s,s^{\prime},q^{2})\,=\,-m_{b}^{2}\,u\,+su-su^{2}-m_{c}^{2}\,v\,+s'v-s\,u\,v\,-s'\,u\,v\,+\,q^{2}\,u\,v\,-s'\,v^{2},
	\end{equation}
	and $\Theta[...]$ denotes the unit step function. We utilize the following definitions:
	\begin{eqnarray}
		&&F\,=\,-1\,+\,u\,+\,v,\notag\\
		&&B\,=\,p^{2}\,-\,m_{b}^{2},\notag\\
		&&B^{\prime}\,=\,p^{\prime2}\,-\,m_{c}^{2}.
	\end{eqnarray}
	
	 \section{THE SUM RULES EXPRESSIONS FOR‌ THE FORM FACTORS}\label{app new}
	The expressions of the sum rules for  the form factors $F_{1},\, F_{2},\, F_{3},\, F_{4},\, G_{1},\, G_{2},\, G_{3}$, and $G_{4}$ used in the calculations are presented in this appendix, 
	 \begin{eqnarray}
	 	F_{1}(q^{2})\,&&=\,\frac{e^{m_{\Sigma_{b}^{*0}}^{2}/M_{1}^{2}}\,e^{m_{\Sigma_{c}^{+}}^{2}/M_{2}^{2}}}{m_{\Sigma_{c}^{+}}\,\lambda_{\Sigma_{b}^{*0}}\,\lambda_{\Sigma_{c}^{+}}}\,\Bigg\lbrace\Bigg[\int_{m_{b}^{2}}^{s_{0}}\mathrm{d}s\int_{m_{c}^{2}}^{s_{0}^{\prime}}\mathrm{d}s^{\prime}\,e^{-s/M_{1}^{2}}\,e^{-s^{\prime}/M_{2}^{2}}\bigg(\rho_{g_{\mu\nu}\,\slashed p\gamma_{5}}^{Pert.}(s,s^{\prime},q^{2})+\rho_{g_{\mu\nu}\,\slashed p\gamma_{5}}^{Dim-3}(s,s^{\prime},q^{2})\bigg)\Bigg]\notag\\&&+\,e^{-m_{b}^{2}/M_{1}^{2}}\,e^{-m_{c}^{2}/M_{2}^{2}}\,\frac{m_{c}\langle \bar{u}\,u\,\rangle\,\langle \bar{d}\,d\,\rangle}{3\sqrt3}\,(\beta-1)\Bigg\rbrace
	 \end{eqnarray}	  
	 \begin{eqnarray}
	 	G_{1}(q^{2})\,&&=\,\frac{e^{m_{\Sigma_{b}^{*0}}^{2}/M_{1}^{2}}\,e^{m_{\Sigma_{c}^{+}}^{2}/M_{2}^{2}}}{m_{\Sigma_{c}^{+}}\,\lambda_{\Sigma_{b}^{*0}}\,\lambda_{\Sigma_{c}^{+}}}\,\Bigg\lbrace\Bigg[\int_{m_{b}^{2}}^{s_{0}}\mathrm{d}s\int_{m_{c}^{2}}^{s_{0}^{\prime}}\mathrm{d}s^{\prime}\,e^{-s/M_{1}^{2}}\,e^{-s^{\prime}/M_{2}^{2}}\bigg(\rho_{g_{\mu\nu}\,\slashed p}^{Pert.}(s,s^{\prime},q^{2})+\rho_{g_{\mu\nu}\,\slashed p}^{Dim-3}(s,s^{\prime},q^{2})\bigg)\Bigg]\notag\\&&+\,e^{-m_{b}^{2}/M_{1}^{2}}\,e^{-m_{c}^{2}/M_{2}^{2}}\,\frac{m_{c}\langle \bar{u}\,u\,\rangle\,\langle \bar{d}\,d\,\rangle}{3\sqrt3}\,(\beta-1)\Bigg\rbrace
	 \end{eqnarray}
	 \begin{eqnarray}
	 	F_{2}(q^{2})\,&&=\,\frac{e^{m_{\Sigma_{b}^{*0}}^{2}/M_{1}^{2}}\,e^{m_{\Sigma_{c}^{+}}^{2}/M_{2}^{2}}m_{\Sigma_{c}^{+}}}{\lambda_{\Sigma_{b}^{*0}}\,\lambda_{\Sigma_{c}^{+}}}\,\Bigg\lbrace\int_{m_{b}^{2}}^{s_{0}}\mathrm{d}s\int_{m_{c}^{2}}^{s_{0}^{\prime}}\mathrm{d}s^{\prime}\,e^{-s/M_{1}^{2}}\,e^{-s^{\prime}/M_{2}^{2}}\bigg(\rho_{p^{\prime}_{\nu}\gamma_{\mu}\slashed p \slashed p^{\prime}\gamma_{5}}^{Pert.}(s,s^{\prime},q^{2})\bigg)\Bigg\rbrace
	 \end{eqnarray}	
	  \begin{eqnarray}
	 	G_{2}(q^{2})\,&&=\,\frac{1}{2}\Bigg\lbrace\frac{e^{m_{\Sigma_{b}^{*0}}^{2}/M_{1}^{2}}\,e^{m_{\Sigma_{c}^{+}}^{2}/M_{2}^{2}}m_{\Sigma_{c}^{+}}}{\lambda_{\Sigma_{b}^{*0}}\,\lambda_{\Sigma_{c}^{+}}}\,\Bigg[\int_{m_{b}^{2}}^{s_{0}}\mathrm{d}s\int_{m_{c}^{2}}^{s_{0}^{\prime}}\mathrm{d}s^{\prime}\,e^{-s/M_{1}^{2}}\,e^{-s^{\prime}/M_{2}^{2}}\bigg(\rho_{p^{\prime}_{\mu}p^{\prime}_{\nu}\slashed p}^{Pert.}(s,s^{\prime},q^{2})\bigg)\Bigg]\notag\\&&+\frac{e^{m_{\Sigma_{b}^{*0}}^{2}/M_{1}^{2}}\,e^{m_{\Sigma_{c}^{+}}^{2}/M_{2}^{2}}m_{\Sigma_{c}^{+}}^{2}}{\lambda_{\Sigma_{b}^{*0}}\,\lambda_{\Sigma_{c}^{+}}}\,\Bigg[\int_{m_{b}^{2}}^{s_{0}}\mathrm{d}s\int_{m_{c}^{2}}^{s_{0}^{\prime}}\mathrm{d}s^{\prime}\,e^{-s/M_{1}^{2}}\,e^{-s^{\prime}/M_{2}^{2}}\bigg(\rho_{p^{\prime}_{\mu}p^{\prime}_{\nu}\slashed p\slashed p^{\prime}}^{Pert.}(s,s^{\prime},q^{2})\bigg)\Bigg]\Bigg\rbrace
	 \end{eqnarray}	 
	  \begin{eqnarray}
	 	F_{3}(q^{2})\,&&=\,\frac{e^{m_{\Sigma_{b}^{*0}}^{2}/M_{1}^{2}}\,e^{m_{\Sigma_{c}^{+}}^{2}/M_{2}^{2}}m_{\Sigma_{c}^{+}}^{2}}{\lambda_{\Sigma_{b}^{*0}}\,\lambda_{\Sigma_{c}^{+}}}\Bigg\lbrace\,\Bigg[-\int_{m_{b}^{2}}^{s_{0}}\mathrm{d}s\int_{m_{c}^{2}}^{s_{0}^{\prime}}\mathrm{d}s^{\prime}\,e^{-s/M_{1}^{2}}\,e^{-s^{\prime}/M_{2}^{2}}\bigg(\rho_{p_{\mu}p^{\prime}_{\nu}\slashed p\slashed p^{\prime}\gamma_{5}}^{Pert.}(s,s^{\prime},q^{2})\bigg)\Bigg]\notag\\&&-\,\Bigg[\int_{m_{b}^{2}}^{s_{0}}\mathrm{d}s\int_{m_{c}^{2}}^{s_{0}^{\prime}}\mathrm{d}s^{\prime}\,e^{-s/M_{1}^{2}}\,e^{-s^{\prime}/M_{2}^{2}}\bigg(\rho_{p^{\prime}_{\mu}p^{\prime}_{\nu}\slashed p\slashed p^{\prime}\gamma_{5}}^{Pert.}(s,s^{\prime},q^{2})\bigg)\Bigg]\Bigg\rbrace
	 \end{eqnarray}	
	  \begin{eqnarray}
	 	G_{3}(q^{2})\,&&=\,\frac{e^{m_{\Sigma_{b}^{*0}}^{2}/M_{1}^{2}}\,e^{m_{\Sigma_{c}^{+}}^{2}/M_{2}^{2}}m_{\Sigma_{c}^{+}}^{2}}{\lambda_{\Sigma_{b}^{*0}}\,\lambda_{\Sigma_{c}^{+}}\,m_{\Sigma_{b}^{*0}}}\Bigg\lbrace\,\Bigg[\int_{m_{b}^{2}}^{s_{0}}\mathrm{d}s\int_{m_{c}^{2}}^{s_{0}^{\prime}}\mathrm{d}s^{\prime}\,e^{-s/M_{1}^{2}}\,e^{-s^{\prime}/M_{2}^{2}}\bigg(\rho_{p_{\mu}p^{\prime}_{\nu}\slashed p^{\prime}}^{Pert.}(s,s^{\prime},q^{2})\bigg)\Bigg]\notag\\&&+\,\Bigg[\int_{m_{b}^{2}}^{s_{0}}\mathrm{d}s\int_{m_{c}^{2}}^{s_{0}^{\prime}}\mathrm{d}s^{\prime}\,e^{-s/M_{1}^{2}}\,e^{-s^{\prime}/M_{2}^{2}}\bigg(\rho_{p^{\prime}_{\mu}p^{\prime}_{\nu}\slashed p^{\prime}}^{Pert.}(s,s^{\prime},q^{2})\bigg)\Bigg]\Bigg\rbrace
	 \end{eqnarray} 
	  \begin{eqnarray}
	 	F_{4}(q^{2})\,&&=\,\frac{e^{m_{\Sigma_{b}^{*0}}^{2}/M_{1}^{2}}\,e^{m_{\Sigma_{c}^{+}}^{2}/M_{2}^{2}}m_{\Sigma_{c}^{+}}^{2}}{\lambda_{\Sigma_{b}^{*0}}\,\lambda_{\Sigma_{c}^{+}}}\,\Bigg\lbrace\int_{m_{b}^{2}}^{s_{0}}\mathrm{d}s\int_{m_{c}^{2}}^{s_{0}^{\prime}}\mathrm{d}s^{\prime}\,e^{-s/M_{1}^{2}}\,e^{-s^{\prime}/M_{2}^{2}}\bigg(\rho_{p^{\prime}_{\mu}p^{\prime}_{\nu}\slashed p \slashed p^{\prime}\gamma_{5}}^{Pert.}(s,s^{\prime},q^{2})\bigg)\Bigg\rbrace
	 \end{eqnarray}	    
	\begin{eqnarray}
		G_{4}(q^{2})\,&&=\,\frac{e^{m_{\Sigma_{b}^{*0}}^{2}/M_{1}^{2}}\,e^{m_{\Sigma_{c}^{+}}^{2}/M_{2}^{2}}m_{\Sigma_{c}^{+}}^{2}}{\lambda_{\Sigma_{b}^{*0}}\,\lambda_{\Sigma_{c}^{+}}}\,\Bigg\lbrace\int_{m_{b}^{2}}^{s_{0}}\mathrm{d}s\int_{m_{c}^{2}}^{s_{0}^{\prime}}\mathrm{d}s^{\prime}\,e^{-s/M_{1}^{2}}\,e^{-s^{\prime}/M_{2}^{2}}\bigg(\rho_{p^{\prime}_{\mu}p^{\prime}_{\nu}\slashed p \slashed p^{\prime}}^{Pert.}(s,s^{\prime},q^{2})\bigg)\Bigg\rbrace
	\end{eqnarray}

\end{document}